\pdfoutput=1

\documentclass[aps,prd,%
onecolumn,%
tightenlines,%
eqsecnum,%
showpacs,%
floats,%
nofootinbib,%
longbibliography,
amsfonts,amssymb,amsmath%
]{revtex4-1}

\usepackage{ifthen}
\makeatletter%
\usepackage{xstring}
\IfSubStr{\@classoptionslist}{preprint}%
{}%
{}%
\makeatother%

\newcommand{\arxiv}{1}

\usepackage{graphicx} 
\usepackage[caption=false]{subfig}             

\graphicspath{{Figures/}}

\usepackage{hyperref}

\usepackage{amsthm}
\usepackage{colordvi} 
\usepackage{enumerate} 

\newcommand{\ket}[1]{\mbox{$| {#1} \rangle$}}
\newcommand{\bracket}[2]{\mbox{$\langle {#1} \!\mid\! {#2} \rangle$}}

\newcommand{\melt}[3]{\mbox{$\langle {#1} | {#2} | {#3} \rangle$}}

\newcommand{\Projsupb}[2]{\mbox{$P^{{#1}}_{\mbox{\scriptsize{${#2}$}}}$}}

\usepackage{dsfont}
\def\Id{\mathds{1}}

\def\Hphys{{\mathcal H}_{\mathrm{phys}}}

\def\d{\textrm{d}}

\def\lp{{l}_{p}}

\def\v{\nu}

\def\Psik{\tilde{\Psi}}

\begin{document}

\preprint{Preprint \#: PI-QG-249}

\title{Consistent probabilities in loop quantum cosmology}

\author{David A.~Craig}
\email[]{E-mail: craigda@lemoyne.edu}
\affiliation{%
Perimeter Institute for Theoretical Physics\\
Waterloo, Ontario, N2L 2Y5, Canada\\
and\\
Department of Chemistry and Physics, Le Moyne College\\
Syracuse, New York, 13214, USA}

\author{Parampreet Singh}
\email[]{E-mail: psingh@phys.lsu.edu}
\affiliation{%
Department of Physics, Louisiana State University\\
Baton Rouge, Louisiana, 70803, USA}

\begin{abstract}

A fundamental issue for any quantum cosmological theory is to specify how
probabilities can be assigned to various quantum events or sequences of events
such as the occurrence of singularities or bounces.  In previous work, we have
demonstrated how this issue can be successfully addressed within the
consistent histories approach to quantum theory for Wheeler-DeWitt-quantized
cosmological models.  In this work, we generalize that analysis to the exactly
solvable loop quantization of a spatially flat, homogeneous and isotropic
cosmology sourced with a massless, minimally coupled scalar field known as
sLQC. We provide an explicit, rigorous and complete decoherent histories
formulation for this model and compute the probabilities for the occurrence of
a quantum bounce vs.\ a singularity.  Using the scalar field as an emergent
internal time, we show for generic states that the probability for a
singularity to occur in this model is zero, and that of a bounce is unity,
complementing earlier studies of the expectation values of the volume and
matter density in this theory.  We also show from the consistent histories
point of view that all states in this model, whether quantum or classical,
achieve arbitrarily large volume in the limit of infinite `past' or `future'
scalar `time', in the sense that the wave function evaluated at any arbitrary
fixed value of the volume vanishes in that limit.
Finally, we briefly discuss certain misconceptions concerning the utility of
the consistent histories approach in these models.

\end{abstract}

%

\maketitle

\section{Introduction}
\label{sec:intro}

When do statements about the behavior of a physical system constitute a
prediction, in the probabilistic sense, of the corresponding quantum theory?
The answer, according to the consistent histories approach to quantum theory
pioneered by Griffiths \cite{griffiths84,*griffiths08}, Omnes
\cite{omnes88a,*omnes88b,*omnes88c,*omnes89,omnes94}, Gell-Mann and Hartle
\cite{GMH90a,*GMH90b,hartle91a,lesH}, Halliwell
\cite{halliwell99,*hallithor01,*hallithor02,*halliwall06,*halliwell09} and others
\cite{hartlemarolf97,CH04,as05}, is when -- and only when -- the quantum
interference between the histories corresponding to those statements vanishes.

A framework of this kind is essential to the quantum theory of gravity applied
to the universe as a whole because the universe is a closed quantum system.
The usual formulation of quantum theory in which measurement by an external
classical observer fixes whether a quantum amplitude determines a quantum
probability is therefore not available \cite{GMH90a,GMH90b,hartle91a,lesH}.
Investigation of real measurement-type interactions
\cite{WZ83,schlosshauer07,GZ05} shows that a key feature of measurements is
that they destroy the interference between alternative outcomes.%
\footnote{The archetypal classroom example of this is the two-slit experiment:
when the experiment is configured in such a way as to gather information
concerning which slit the particle passed through, the interference pattern is
destroyed, and meaningful probabilities -- in the sense that the usual
probability sum rules are satisfied -- may be assigned to the alternative
histories \{(upper slit,$y$),(lower slit,$y$)\}, where $y$ is the position of
arrival of the electron on the screen.  If information concerning the slit is
not gathered, there is interference between the alternative histories, and no
physically meaningful probability can be assigned to the slit through which
the particle passed before arriving at position $y$.  Quantum theory simply
has no prediction concerning such histories.  More pointedly, in quantum
theory \emph{there can be no logically consistent prediction} for these
particular histories in this experimental configuration.
} %
The consistent or decoherent histories approach to quantum theory formalizes
this observation by supplying an objective, observer-independent measure of
quantum interference between alternative histories called the \emph{decoherence
functional}.  The decoherence functional, constructed from the system's
quantum state, both measures the interference between histories in a
complete set of alternative possibilities, and, when that interference
vanishes between all members of such a set, determines the probabilities of
each such history.  This framework reproduces the ordinary quantum quantum
mechanics of measured subsystems in situations to which it applies, but
generalizes it to situations in which it does not, such as when applying
quantum theory to a closed system such as the universe as a whole.

In previous work \cite{CS10a,CS10b,CS10c}, we have developed the consistent
histories framework for a model quantum gravitational system, a Wheeler-DeWitt
quantization \cite{aps} of a spatially flat Friedmann-Robertson-Walker (FRW)
cosmology sourced by a free, massless, minimally-coupled scalar field.  In
this paper, we give the consistent histories formulation of the corresponding
model \cite{aps,aps:improved,acs:slqc} in loop quantum cosmology (LQC).  (See
Ref.\ \cite{ashsingh11} for a review of LQC.) A key prediction of LQC is the
existence of a bounce of the physical volume (or the scale factor) of the
universe when the energy density of the matter content (in the present case,
the scalar field $\phi$) reaches a universal maximum $\rho_{\mathrm{max}} =
0.41\rho_{\mathrm{Planck}}$ in the isotropic models.%
\footnote{The existence of a bounce in LQC has also been demonstrated for
anisotropic models, however due to contributions from the anisotropic shear to
the spacetime curvature, the bounce of the mean scale factor is not
accompanied by a saturation of the energy density.  The energy density and the
shear scalar are however still bounded above.  (See Ref.\ \cite{guptsingh12a}
for more discussion in different Bianchi models.)
} %
The existence of a bounce was first obtained for the model under consideration
\cite{aps_letter,aps,aps:improved}, and since then has been confirmed for a
variety of matter models, using sophisticated numerical
simulations.%
\footnote{For a summary of results on the bounce in various models in LQC, see
Ref.\ \cite{ashsingh11}.  A review of numerical techniques, including the way
comparisons with the classical theory are made, is available in Ref.\
\cite{singh12a}.
} %
These numerical simulations show that semi-classical states peaked at late
times on classical expanding trajectories, bounce in the backward evolution
(in `internal time' $\phi$) to a classical contracting branch.  Since the
inner product, physical Hilbert space and a set of Dirac observables are
completely known, the detailed physics can be extracted and reliable predictions
can be made.
Interestingly, the spatially flat isotropic model with a massless scalar field
can be solved exactly in LQC \cite{acs:slqc}.  This model, dubbed sLQC, serves
as an important robustness check of various predictions in loop quantum
cosmology.  In particular, it has been shown that the bounce occurs for all
the states in the physical Hilbert space, and the energy density is bounded
above by the same universal maximum $\rho_{\mathrm{max}}$ 
\cite{acs:slqc,dac13a}.

All of these studies, though, address in practice only questions concerning
\emph{individual} quantum events, for example, the density or volume (of a
fiducial spatial cell of the universe) at a given value of internal time.
However, as discussed in detail in Refs.\ \cite{CS10a,CS10b,CS10c},
conclusions drawn from such individual quantum events can be in certain
situations badly misleading as a guide to probabilities for \emph{sequences}
of quantum occurrences -- \emph{histories} of the universe -- precisely the
kind of physical questions in which we are most interested in the context of
the physics of cosmological history.  The question is, when does the
\emph{amplitude} for a sequence of quantum events correspond to the
\emph{probability} for that particular history?  The answer is, when, and only
when, the interference between the alternative histories vanishes -- just as
in the two-slit experiment -- as determined by the system's decoherence
functional.

In this paper we construct the decoherence functional for sLQC and employ it
to study quantum histories of physical observables, concentrating on
the physical volume of the fiducial cell.
We examine both semiclassical and generic quantum states.  
We work within a complete predictive framework for the quantum mechanics of
\emph{history} to study the physics of the quantum bounce, showing that the
corresponding quantum histories decohere, and that the probability of a
cosmological bounce in these models is unity for generic quantum states (not
just semiclassical ones).  This stands in stark contrast to the predictions
for the Wheeler-DeWitt quantization of the same model, which is shown in
Refs.\ \cite{acs:slqc,CS10c} to be certain to be \emph{singular} for generic
quantum states.

We close this introduction with a note on the role played in quantum cosmology
by larger issues in the interpretation of quantum mechanics.  It is perhaps an
understatement to observe that the philosophical challenges presented by the
effort to apply quantum theory to closed systems -- particularly, the universe
as a whole -- do not end with questions of consistency of histories or
decoherence.  A fundamental challenge to the program is to offer a coherent
account of the physical meaning of the probabilities at which one consistently
arrives \cite{hartle91a,sorkin94,sorkin97a,MW10}.  This profound question is
not the subject of this paper.  Indeed, there is little agreement on the
``true'' nature of probability even in classical physics, never mind quantum
mechanics more broadly \cite{ballentine01,jaynes03} or the quantum theory of
closed systems in particular.  
Here we adopt the pragmatic attitude fairly typical in physics.  When multiple
instantiations of the ``same'' physical system are available, probability is
interpreted through ``for all practical purposes'' operational definitions
based on relative frequencies of outcomes \cite{hartle68,caves-schack05a}.  For
single systems (such as the whole universe), a frequentist interpretation
is not so easily accessible.%
\footnote{For some recent discussion, see Ref.\ \cite{MW10}.
} %
Even though we do not shy away from writing down probabilities in this paper,
we recognize the interpretational challenges and therefore concentrate
particularly on a class of quantum predictions for which the interpretation of
probabilities might be hoped to be less controversial: those which are
\emph{certain} i.e.\ have probabilities equal to 0 or 1 -- or very close
thereto \cite{hartle88a,hartle91a,sorkin94,sorkin97a}.  

The plan of the paper is as follows.  In Sec.\ II we briefly summarize the
framework of loop quantum cosmology and discuss the quantization of sLQC.
Starting from the classical theory formulated in Ashtekar variables, we show
the way inner product, physical Hilbert space and Dirac observables are
constructed, and an evolution equation in the emergent `internal time' $\phi$
is obtained.  In Sec.\ III, we summarize generalized decoherent (or
consistent) histories quantum theory in the context in sLQC, by rewriting the
standard approach in proper time in terms of the internal time $\phi$.  We
describe the construction of the generalized quantum theory for sLQC,
including definitions of its class operators (histories), branch wave
functions, and decoherence functional.  (More details of the classical theory
of the model considered and the standard consistent histories approach can be
found in our previous work \cite{CS10a,CS10b,CS10c}.)
In Sec.\ \ref{sec:apps} we apply these constructions to quantum predictions
concerning histories of the cosmological volume by using some important
properties of the eigenfunctions of the quantum Hamiltonian constraint derived
recently \cite{dac13a}.  We first introduce class operators for the volume
observable, and discuss the way probabilities can be computed for histories
involving single and multiple instants of internal time $\phi$.  We evaluate
the probability for occurence of a quantum bounce for semi-classical states,
as well as generic states.  We show that the probability of occurence of a
bounce in sLQC turns out to be unity for \emph{all} states in the theory.
Sec.\ \ref{sec:discuss} closes with some discussion.

\section{Loop quantization of flat, homogeneous and isotropic cosmology}
\label{sec:sLQC}

In this section, we briefly outline the quantization of a spatially flat,
homogeneous and isotropic spacetime in loop quantum cosmology.%
\footnote{For various details, see Refs.\ \cite{aps,aps:improved,acs:slqc,ashsingh11}.
} %
A complete loop quantization of this model sourced with a massless, minimally
coupled scalar field $\phi$ was first provided in Refs.\
\cite{aps_letter,aps,aps:improved}, and the model was demonstrated in Ref.\
\cite{acs:slqc} to be exactly solvable in the ``harmonic gauge'' $N = a(t)^3$,
where $a(t)$ denotes the scale factor of the universe described by the
Friedmann-Lema\^{i}tre-Robertson-Walker metric
\begin{equation}
g_{ab} = -N^2\d t_a \d t_b + a^2(t) \mathring{q}_{ab}.
\end{equation}
Here $\mathring{q}_{ab}$ is a flat fiducial metric on the spatial slices
$\Sigma$.  In loop quantum cosmology the quantization procedure parallels that
of loop quantum gravity (LQG).  The gravitational phase space variables in
loop quantum cosmology are the symmetric connection $c$ and its conjugate
triad $p$, obtained by a symmetry reduction of the gravitational phase space
variables in LQG, the Ashtekar-Barbero SU(2) connection $A^i_a$, and the
densitized triad $E^a_i$.  These are related by
\begin{equation}
A^i_a \, = \, c \, V_o^{-1/3} \mathring{\omega}^i_a, 
  ~~~ E^a_i = p \sqrt{\mathring{q}}\, V_o^{-2/3} \mathring{e}^a_i ~.
\end{equation}
Here $V_o$ denotes the volume with respect to $\mathring{q}_{ab}$ of a
fiducial cell introduced in order to define a symplectic structure on
$\Sigma$,%
\footnote{This choice is necessary if the topology of the manifold is
non-compact (in the present case, $\mathbb{R}^3$), as chosen in this analysis.
The results obtained are independent of this choice, and are unaffected if we
choose a compact topology (in the present case, $\mathbb{T}^3$).  See Ref.\
\cite{ashsingh11} for a brief discussion of this point.
} %
and $\mathring{e}^a_i$ and $\mathring{\omega}^i_a$ respectively denote a
fiducial triad and co-triad compatible with the fiducial metric.  (In these
variables the physical volume of the fiducial cell is $V=a^3V_o=|p|^{3/2}$.)
For the massless scalar field model, the matter phase space variables are
$\phi$ and its conjugate momentum $p_\phi$.  In terms of these phase space
variables, the classical Hamiltonian constraint ${\cal C}_{cl}$ can be written
as
\begin{equation}\label{eq:classicalH}
{\cal C}_{cl} = - 3 \pi G \hbar^2 b^2 \nu^2 + p_\phi^2,
\end{equation}
where $b$ and $\nu$ are related to $c$ and $p$ by
\begin{equation}
b \, = \, \frac{c}{|p|^{1/2}}, ~~~~ \nu = \frac{|p|^{3/2}}{2 \pi \gamma l_p^2} ~.
\end{equation}
Here $l_p=\sqrt{G\hbar}$ is the Planck length.  (We have set $c=1$.)  Note
that $\nu$, though a measure of the physical volume of the fiducial cell,
has dimensions of length.  The modulus sign arises due to the two physically
equivalent orientations of the triad.  We will choose the orientation to be
positive without any loss of generality.  

Hamilton's equations for Eq.\ (\ref{eq:classicalH}) yield the classical
trajectories via the Poisson brackets $\{b,\nu\} = 2 \hbar^{-1}$ and $\{\phi,
p_\phi\} = 1$.  These yield $p_\phi = V_o a^3 \dot \phi$ as a constant of
motion, and relate $\phi$ and $\nu$ by
\begin{equation}\label{eq:classtraj}
\phi = 
 \pm \frac{1}{\sqrt{12 \pi G}} \, \ln \left|\frac{\nu}{\nu_o}\right| + \phi_o~,
\end{equation}
where $\nu_o$ and $\phi_o$ are constants of integration.  In the classical
theory, for $\nu \geq 0$ and regarding $\phi$ as an emergent internal physical
`clock', there exist two disjoint solutions, one expanding and the other
contracting, with a fixed value of $p_\phi$.  In the limit $\phi \rightarrow
-\infty$ the expanding branch encounters a big bang singularity in the past
evolution, whereas in the limit $\phi \rightarrow \infty$ the contracting
branch encounters a big crunch singularity in the future evolution.  These
singularities are reached in a finite proper time, and {\it{all}} the
classical solutions are singular.

We now summarize the quantization procedure for this model in loop quantum
cosmology in brief.  As in LQG, the fundamental variables for quantization of
the gravitational sector are the holonomies of the connection and the fluxes
of the triads.  Due to spatial homogeneity, the fluxes turn out to be
proportional to the triads themselves \cite{abl03}, whereas the holonomies of
the connection, along straight edges labelled by $\mu$, are given by
\begin{equation}
h_k^{(\mu)} = \cos \left(\frac{\mu c}{2} \right) \mathbb{I} 
     - 2 i \sin  \left(\frac{\mu c}{2} \right) \frac{\sigma_k}{2} ~,
\end{equation}
where the $\sigma_k$ are the Pauli spin matrices.  The matrix elements of the
holonomies generate an algebra of almost periodic functions of the connection,
the representation of which, found via the Gel'fand-Naimark-Segal contruction,
supplies the kinematical Hilbert space.  It turns out that even at the
kinematical level, the quantization of this model in LQC is strikingly
different from that of the Wheeler-DeWitt theory.  The gravitational sector of
the kinematical Hilbert space in loop quantum cosmology is ${\cal
H}_{\mathrm{grav}}^{\mathrm{(kin)}} = L^2(\mathbb{R}_{\mathrm{Bohr}}, d
\mu_{\mathrm{Bohr}})$ where $\mathbb{R}_{\mathrm{Bohr}}$ is the Bohr
compactification of the real line, and $\mu_{\mathrm{Bohr}}$ is the Haar
measure on it.  In contrast, the kinematical Hilbert space in the
Wheeler-DeWitt theory is $L^2(\mathbb{R}, d c)$.  Unlike the Wheeler-DeWitt
theory, a generic state in the kinematical Hilbert space of LQC can be
expressed as a countable sum of orthonormal eigenfunctions (matrix
elements of holonomies).

The matrix elements of the holonomies act on states in the volume (or the
triad) representation as translations.  If $|\nu\rangle$ denotes the
eigenstates of the volume operator, which has the action $\hat V |\nu \rangle
= 2 \pi \gamma \lp^2 |\nu| |\nu\rangle$, then elements of the holonomies act
as $\widehat{\exp(i \lambda b)} |\nu\rangle = |\nu - 2 \lambda \rangle$.  Here
$\lambda$ is a parameter determined by the underlying quantum geometry, and is
given by $\lambda^2 = 4 \sqrt{3} \pi \gamma \lp^2$ \cite{awe09a}.  A
consequence is that the action of the Hamiltonian constraint operator,
expressed in terms of the holonomies, on the states in the volume
representation does not lead to a differential equation, but rather to a
difference equation in which the discreteness scale is determined by the
parameter $\lambda$.  For the total Hamiltonian constraint $\hat
C_{\mathrm{grav}} + 16 \pi G \hat C_{\mathrm{matt}} \approx 0$, the resulting
difference equation is given by
\begin{subequations}
\begin{eqnarray}
\Theta\Psi(\v,\phi)  & := &  -\frac{3\pi G}{4\lambda^2} \left\{
\sqrt{|\v(\v+4\lambda)|} |\v+2\lambda| \Psi(\v+4\lambda,\phi)
- 2 \v^2 \Psi(\v,\phi)  \right. \nonumber\\
  & \qquad &  \qquad \qquad \qquad \qquad \qquad 
      \left. + \sqrt{|\v(\v-4\lambda)|} |\v-2\lambda| \Psi(\v-4\lambda,\phi) \right\} 
\label{eq:theta-a}\\  
& = & - \partial_\phi^2 \Psi(\nu,\phi)~,
\label{eq:theta-b}
\end{eqnarray}
\label{eq:theta}%
\end{subequations}
where the gravitational part of the constraint $\Theta$ is a self-adjoint,
positive definite operator.%
\footnote{This expression is different from what is found in Refs.\
\cite{aps:improved,acs:slqc} because we are using states that carry an
additional factor of $\sqrt{\lambda/|\v|}$ relative to those states in order
to simplify the form of the inner product, Eq.\ (\ref{eq:ip}).  Compare, for
example, Refs.\ \cite{ach10a,ach10b}.  
\label{foot:statenorm}
} %
The similarity of this equation to the Klein-Gordon equation is compelling.
Since $\phi$ is monotonic, it may be treated as an emergent internal time.
Solutions of the constraint equation can then be divided into orthogonal,
physically equivalent positive and negative frequency subspaces.  As in the
Klein-Gordon theory, it suffices to consider only one of these subspaces to
extract physics.  We consider states lying in the positive frequency subspace,
satisfying
\begin{equation}\label{eq:evolutioneq_pf}
-i \partial_\phi \Psi(\nu,\phi) = \sqrt{\Theta} \Psi(\nu,\phi)~,
\end{equation}
which are normalized with respect to the inner product 
\begin{equation}
\bracket{\Psi}{\Phi} = \sum_{\v=4\lambda n} \Psi(\v,\phi_o)^*\Phi(\v,\phi_o) ~.
\label{eq:ip}
\end{equation}
Note the inner product so defined is \emph{independent} of the choice of
$\phi_o$.  Eq.\ (\ref{eq:ip}) defines the Hilbert space of sLQC.

Physical states have a support on the lattices $\nu = (4 n \pm
\epsilon)\lambda$, with $n \in \mathbb{Z}$ and $\epsilon \in [0,4)$.  Thus,
there is super-selection among lattices with different $\epsilon$.  In this
manuscript, we will focus on the $\epsilon =0$ sector, which allows the states
to have support on zero volume -- the big bang singularity in the classical
theory. 

An additional requirement on the physical states $\Psi(\nu,\phi)$ arises by
noting that in the absence of fermions, physics should be independent of the
orientation of the triad.  We can thus choose physical states to be symmetric
under this change, which therefore satisfy $\Psi(\v,\phi)=\Psi(-\v,\phi)$.  
Because of the symmetry of the physical states and observables under changes 
of orientation of the triad, we will as applicable treat $\nu$ as positive 
in the sequel.

In order to extract physics, we introduce a set of Dirac observables.  These
are the volume of the fiducial cell at time $\phi^*$, and the conjugate momentum
$p_\phi$, which have the following action (consistent with the inner product):
\begin{equation}
\hat V|_{\phi^*} \Psi(\nu,\phi) = 2 \pi \gamma \lp^2 e^{i \sqrt{\Theta} (\phi - \phi^*)} |\nu| \Psi(\nu,\phi^*), 
  ~~\hat p_\phi\Psi(\nu,\phi) = \hbar \sqrt{\Theta} \Psi(\nu,\phi) ~.
\end{equation}
Using these observables, it is straightforward to also introduce an energy
density observable, which turns out to have expectation values bounded above
by a critical density $\rho_{\mathrm{max}}$ for all the states in the physical
Hilbert space \cite{acs:slqc,dac13a}.  Analysis of these observables in sLQC,
in confirmation with the earlier results in LQC obtained using numerical
simulations \cite{aps_letter,aps,aps:improved}, show that the expectation
value of the volume observable has a minimum which is reached when the energy
density reaches its maximum value.  This is the quantum bounce in sLQC. Our
goal is now to understand the occurrence of a bounce in sLQC using the
consistent histories approach, which is addressed in the following.

\section{Consistent histories formulation of sLQC}

In this section we apply the ideas of the consistent histories approach to
quantum mechanics (also known as `generalized quantum theory' a la Hartle
\cite{hartle91a,lesH}) to the sLQC model discussed in Sec.\ \ref{sec:sLQC}.
The formalism will then be used in Sec.\ \ref{sec:apps}, to make
quantum-mechanically consistent predictions concerning the behavior of the
physical universe by employing the decoherence functional to measure the
quantum interference between possible alternative histories.

Our definitions will naturally directly mirror those for the Wheeler-DeWitt
quantization of the same model \cite{CS10a,CS10b,CS10c}, facilitating easy
comparison of the sometimes divergent predictions of the two models.
Moreover, as noted, the construction of class operators, branch wave
functions, and the decoherence functional for sLQC precisely mirrors that of
the Wheeler-DeWitt theory.  In this section, therefore, we restrict ourselves
to a concise summary of the definitions and main formul\ae, referring the
reader to Ref.\ \cite{CS10c} for a more in depth discussion and commentary.

The three essential ingredients of a generalized quantum theory are: (i) The
\emph{fine-grained histories}, the most refined descriptions of a system it is
possible to give.  (These might be individual paths in a path integral
formulation of the theory, for example.)  (ii) The \emph{coarse-grained
histories}, a specification of the allowed partitions of the fine-grained
histories into physically meaningful subsets.  (Only diffeomorphism invariant
partitions might be allowed in a covariant quantization of gravity, for
example.)  Since most all physical predictions concern highly coarse-grained
descriptions of the universe, it is the coarse-grained histories which
correspond to physically meaningful questions, and for which quantum theory
must be able to determine probabilities -- and indeed, if those probabilities
are meaningful at all.  (iii) The \emph{decoherence functional} provides an
objective, observer-independent measure of the quantum interference between
alternative coarse-grained histories of a system.  When that interference
vanishes among all members of a coarse-grained family, that set is said to
``decohere'', or to ``be consistent''.  In that case, and in that case only,
does the decoherence functional assign logically consistent probabilities --
in the sense that probability sum rules are satisfied -- to the members of
each consistent set of histories.

Any specific implementation of a generalized quantum theory must realize these
elements in a coherent and mathematically consistent way.
In formulations of quantum theory in Hilbert space, fine-grained histories can
be specified by (for example) time-ordered products of Heisenberg projections
onto eigenstates of physical observables, representing the history in which
the system assumes those particular values of those particular observables at
those particular times.  Coarse-grained histories are represented by sums of
such fine-grained histories.  ``Branch wave functions'' corresponding to the
state of a system that has followed a particular coarse-grained history are
defined by the action of these history (or ``class'') operators on the quantum
state.  The decoherence functional, which measures the interference between
alternative histories, and also the probabilities of histories in consistent
or decoherent families as determined by the absence of such interference, can
be defined by the physical inner product between branch wave functions.

In the consistent histories approach to ordinary non-relativistic quantum
theory, histories are defined using coordinate time $t$.  As discussed in the
previous section, in sLQC, the role of time is naturally played by the
massless scalar field $\phi$.  Indeed, using Eq.\ (\ref{eq:evolutioneq_pf}) we see
that the states $|\Psi\rangle$ evolve unitarily in $\phi$,
\begin{equation}\label{eq:proppsi}
\Psi(\nu,\phi) = e^{i \sqrt{\Theta} (\phi- \phi_o)} \Psi(\nu,\phi_o) =: U(\phi - \phi_o)  \Psi(\nu,\phi_o)~.
\end{equation}
$U(\phi)$ is thus the propagator for evolution in `time' $\phi$.  Using $\phi$
as the internal time, we can define Heisenberg projections in analogy with
non-relativistic quantum mechanics and obtain class operators, branch
wave functions, and a decoherence functional.%
\footnote{For comparison, Ref.\ \cite{CH04} constructed the decoherence
functional, following earlier work \cite{hartle91a,lesH}, for a path integral
quantization of closed type A minisuperspace models.  The decoherence
functional for spin foam models \cite{CS12c} and other quantum gravitational
models without an internal time fits naturally into this more general
framework.
} %
In this set up, the class operators provide predictions concerning histories
of values of the Dirac observables.  Our strategy here directly parallels the
one we followed for the quantization of the Wheeler-DeWitt model with a
massless scalar field \cite{CS10c}.

\subsection{Class operators}
\label{sec:LQCclassops}

Class operators correspond to the physical questions that may be asked of a
given system.  All such questions come in exclusive, exhaustive sets -- at the
most coarse-grained level, simply ``Does the universe have property ${\mathcal
P}$, or not?''  The sum of all the class operators in such an exclusive,
exhaustive set must therefore be, in effect, the identity, up to an overall
unitary factor. Homogeneous class operators describe possible sequences of (ranges of) values
of observable quantities, with sums of them 
corresponding to coarse-grainings thereof.  We will often refer to class
operators simply as ``histories''.

In quantum cosmology relevant physical questions include ``What is the
physical volume of the fiducial cell when the scalar field has value
$\phi^*$?''  ``Does the volume of the cell ever drop below a particular value,
let us say $\nu^*$?''  ``Is the momentum of the scalar field conserved during
evolution?''  Does the density exceed $\rho^*$?''  -- and so forth.
In the present model, which possesses a physical clock -- the monotonic
(unitary) internal time supplied by the scalar field $\phi$ -- class operators
for questions of this kind may be constructed similarly to those of
non-relativistic quantum theory, in which fine-grained class operators
correspond to predictions concerning the values of physical observables at
given moments of time.  From a physical point of view, in quantum cosmology,
class operators constructed in a similar manner correspond to physical
questions concerning the \emph{correlation} between values of various
observable quantities and the value of the scalar field.  It is no surprise,
then, that class operators of this type naturally correspond to predictions
concerning the values of relational observables, as noted in Sec.\
\ref{sec:LQCrelnl}.%
\footnote{Relational questions of this form are certainly not the only
physical questions in which one might be interested, or for which class
operators may be constructed -- either in ordinary quantum mechanics
\cite{yt91a,*yt91b,*yt92,*y92,*y96,*lesH,*whelan94,*micanek96} or in quantum
cosmology
\cite{hartlemarolf97,*hallithor01,*hallithor02,*CH04,*halliwall06,*as05,*halliwell09}.
However, they are sufficient for the physical problems considered in this
paper.
} %
Stated in this way, it is clear that the interpretation of the scalar field
$\phi$ as a background physical clock is an inessential, if useful, feature
of this particular model.

In sLQC, we have states $\ket{\Psi}$ with a unitary evolution in $\phi$ given
by Eq.\ (\ref{eq:proppsi}).  As noted, among the physical questions of
interest are the values of volume and scalar momentum at given values of
$\phi$.  To extract physical predictions concerning quantities of this kind,
we proceed as in ordinary quantum theory.  We consider a family of observables
$A^\alpha$, labelled by index $\alpha$, with eigenvalues $ a_k^\alpha$ in the
physical Hilbert space ${\cal H}_{\rm{phys}}$ of sLQC. We denote the ranges of
eigenvalues as $\Delta a_k^\alpha$.  Projections onto the corresponding
eigensubspaces will be denoted  $P_{a_k}^\alpha$ and $P_{\Delta
a_k}^\alpha$, respectively.  For a given choice of observable $A^{\alpha_i}$
at each time $\phi_i$, an exclusive, exhaustive set of fine-grained histories
in sLQC may be regarded as the set of sequences of eigenvalues
$\{h\}=\{(a^{\alpha_1}_{k_1},a^{\alpha_2}_{k_2},\dots,a^{\alpha_n}_{k_n})\}$,
corresponding to the family of histories in which observable $A^{\alpha_i}$
has value $a^{\alpha_i}_{k_i}$ at time $t_i$.  (Each $k_i$ for fixed $i$
therefore runs over the full range of the eigenvalues $a^{\alpha_i}_{k_i}$.)
A different choice of observables $(\alpha_1,\alpha_2,\dots,\alpha_n)$ leads
to different exclusive, exhaustive families of histories $\{ h \}$.  Using the
propagator
 \begin{equation}
U(\phi_i-\phi_j) = e^{i \sqrt{\Theta} (\phi_i-\phi_j)}
\label{eq:propdefqm}
\end{equation}
we define ``Heisenberg projections''
\begin{equation}
\Projsupb{\alpha}{\Delta a^{\alpha}_k}(\phi) =
  U^{\dagger}(\phi-\phi_o) \Projsupb{\alpha}{\Delta a^{\alpha}_k} U(\phi-\phi_o),
\label{eq:HprojDqc}
\end{equation}
where
$\phi_o$ is a value of the scalar field at which the quantum state is defined.%
\footnote{Because the evolution in $\phi$ is unitary, this value is completely
arbitrary and may be adjusted as necessary to remain outside the region of
coarse-grainings of physical interest.
} %
The fine-grained history $h$ may then be conveniently represented by the
class operator%
\footnote{Class operators constructed in this way are subject to the quantum
Zeno effect if spaced too closely in `time''
\cite{halliwell09,halliyear09a,halliyear10a}, and a more subtle definition
may be required.  We do not address this question in this paper, though we may
return to it in a later work.
} %
\begin{subequations}
\begin{eqnarray}
C_h &=& \Projsupb{\alpha_1}{a_{k_1}}(\phi_1)
        \Projsupb{\alpha_2}{a_{k_2}}(\phi_2) \cdots
        \Projsupb{\alpha_n}{a_{k_n}}(\phi_n)
\label{eq:classopdefqm-a}  \\
    &=& U(\phi_o-\phi_1)\Projsupb{\alpha_1}{a_{k_1}}
        U(\phi_1-\phi_2)\Projsupb{\alpha_2}{a_{k_2}} \cdots
        U(\phi_{n-1}-\phi_n)\Projsupb{\alpha_n}{a_{k_n}}U(\phi_n-\phi_o).
\label{eq:classopdefqm-b}
\end{eqnarray}
\label{eq:classopdefqm}%
\end{subequations}%
Since  $\sum_{k}\Projsupb{\alpha}{a_k}=\Id$ for each observable
$\alpha$, the class operator $C_h$ satisfies
\begin{equation}\label{eq:classopsumqm}
\sum_h C_h \, = \, \sum_{k_1}\sum_{k_2}\cdots \sum_{k_n} \,
                     \Projsupb{\alpha_1}{a_{k_1}}(\phi_1)
                     \Projsupb{\alpha_2}{a_{k_2}}(\phi_2) \cdots
                     \Projsupb{\alpha_n}{a_{k_n}}(\phi_n) \, = \, \Id ,
\end{equation}
corresponding to the fact that the set of fine-grained histories $\{ h \}$
represents a mutually exclusive, collectively exhaustive description of the
possible fine-grained histories in sLQC.

The coarse-grained history
\begin{equation}
h =
(\Delta a^{\alpha_1}_{k_1},\Delta a^{\alpha_2}_{k_2},\dots,\Delta a^{\alpha_n}_{k_n}),
\label{eq:historyqc}
\end{equation}
in which the variable $\alpha_1$ takes values in $\Delta a^{\alpha_1}_{k_1}$
at $\phi=\phi_1$, variable $\alpha_2$ takes values in $\Delta a^{\alpha_2}_{k_2}$
at $\phi=\phi_2$, and so on, then has the class operator
\begin{equation}\label{eq:classopdefqc}
C_h = \Projsupb{\alpha_1}{\Delta a_{k_1}}(\phi_1)
        \Projsupb{\alpha_2}{\Delta a_{k_2}}(\phi_2) \cdots
        \Projsupb{\alpha_n}{\Delta a_{k_n}}(\phi_n),
\end{equation}
where we suppress the superscripts on the eigenvalue ranges to minimize
notational clutter.  It is straightforward to see that the class operators for
the coarse grained histories satisfy $\sum_h C_h = \Id$,
the identity on the physical Hilbert space $\Hphys$.

\subsection{Branch wave functions}
\label{sec:LQCbwfs}

Class operators capture the physical questions that may be asked of a system,
as specified by an exclusive, exhaustive set of histories $\{ h \}$.  The
amplitude for a quantum state $\ket{\Psi}$ specified at $\phi=\phi_o$ to
``follow'' one of the histories $h$ -- i.e.\ for the universe to have the
properties described by $h$ -- is given by the branch wave function
$\ket{\Psi_h}$.%
\footnote{Because of the unitary evolution of Eq.\ (\ref{eq:proppsi}), the
choice of $\phi_o$ has no special physical significance.   Indeed, from the
point of view of the consistent histories framework, the quantum state of the
system is most naturally viewed as a ``timeless'' property of the system,
with the more familiar notion of the ``state at a moment of time'' being
recovered, under appropriate circumstances, in the branch wave function
\cite{hartle91a,lesH}.
} %

The branch wave function for a history $h$ in the physical Hilbert space of
sLQC is defined in a manner parallel to non-relativistic quantum mechanics.
Defining
\begin{equation}
C_h(\phi) \, = \, C_h \, \cdot \, U^\dagger(\phi - \phi_o) ~,
\end{equation}
the branch wave function is given by
\begin{subequations}
\begin{eqnarray}
\ket{\Psi_h(\phi)} &=& \noindent U(\phi-\phi_o)C_h^{\dagger}\ket{\Psi} 
\label{eq:bwfdefqc-a}\\
&=& U(\phi-\phi_n)\Projsupb{\alpha_n}{a_{k_n}}
        U(\phi_{n}-\phi_{n-1})\cdots
        U(\phi_2-\phi_1)\Projsupb{\alpha_1}{a_{k_1}}
        U(\phi_1-\phi_o)\ket{\psi}~.
\label{eq:bwfdefqc-b}
\end{eqnarray}
\label{eq:bwfdefqc}%
\end{subequations}%
This branch wave function is, by construction, a solution to the quantum
constraint everywhere.  The propagator $U$ simply evolves the branch wave
function to any convenient choice of $\phi$.  All inner products will of
course be independent of this choice.  The projections implement, in the
standard Copenhagen interpretation, ``wave function collapse''.  From the
consistent histories point of view, however, the branch wave function is
viewed merely as an amplitude
from which one may ultimately construct the probabilities of individual
histories -- the likelihoods that the universe possesses these particular
sequences of physical properties.  In particular, the ``collapse'' is not to
be regarded as a physical process in this framework.

\subsection{The decoherence functional}
\label{sec:LQCd}

Given a complete exclusive, exhaustive set of histories $\{ h \}$ and a
quantum state $\ket{\Psi}$ in the physical Hilbert space, the decoherence
functional measures the interference among the branch wave functions
$\ket{\Psi_h}$, and, if that interference vanishes, determines also the
probabilities of each of the $\ket{\Psi_h}$ -- in other words, the probability
that a universe in the state $\ket{\Psi}$ has the physical properties
described by the history $h$.  If the interference does not vanish, then
quantum theory can make no predictions concerning the particular set of
physical questions $\{ h \}$, in just the same way the question of which
slit a particle passed through cannot be coherently analyzed when it is not
recorded.

The decoherence functional in non-relativistic quantum mechanics is defined by
\begin{equation}
d(h,h') =  \bracket{\Psi_{h'}}{\Psi_{h}},
\label{eq:dfdefqc}\\
\end{equation}
and from Eq.\ (\ref{eq:classopsumqm}) is normalized, $\sum_{h, h'} d(h, h')
\, = 1$.  In quantum cosmology the decoherence functional may be constructed
from the branch wave functions in essentially
the same manner \cite{CH04}.%
\footnote{The generalization of this definition to mixed states is also given
in \cite{CH04}.
} %

``Decoherent'' or ``consistent'' sets of histories are by definition 
exclusive, exhaustive sets of histories $\{ h \}$ which satisfy
\begin{equation}
d(h,h') = p(h)\, \delta_{h'h}
\label{eq:mediumdecoh}
\end{equation}
among all members of the set.  Here $p(h)$ is the probability for the history
$h$.  The physical meaning of this expression is the following.  When
interference between all the members of an exclusive, exhaustive set of
coarse-grained histories $\{ h \}$ vanishes,
$ d(h,h') = 0$ for $h\neq h'$,
that set of histories is said to decohere, or be consistent.  In such sets,
the probabilities of the individual histories are then simply the diagonal
elements of the decoherence functional,
$p(h) = d(h,h)$.
It is easily verified that this is simply the standard L\"{u}ders-von Neumann
formula for probabilities of sequences of outcomes in ordinary quantum
theory, when such probabilities may be defined -- typically in measurement
situations or interactions with an external ``environment'' that leads to
decoherence in the now-conventional sense \cite{schlosshauer07}.
In the framework of generalized (decoherent histories) quantum theory,
however, no external notion of observers or measurement or environment is
required.  
It is the objective, observer independent criterion of Eq.\
(\ref{eq:mediumdecoh}) that determines when probabilities may be defined, and
which ensures these probabilities are meaningful in the sense that probability
sum rules are obeyed when histories are coarse-grained:
$p(h_1+h_2) = p(h_1) + p(h_2)$,
with $\sum_h p(h) = 1$.  If histories do not decohere, then the diagonal
elements of the decoherence functional do not sum to unity and cannot
therefore be interpreted as probabilities.  For such families of histories,
quantum theory is silent: it simply has no logically consistent predictions at
all.  

Note that as constructed, the decoherence functional for sLQC involves an
inner product of branch wave functions on a minusuperspace slice of fixed
$\phi$.
The unitary evolution in $\phi$, and the fact that the branch wave functions
$\ket{\Psi_h(\phi)}$ are by construction everywhere solutions to the
quantum constraint, makes the specific choice of $\phi$ irrelevant in the
definition of the branch wave functions and decoherence functional, and
therefore may be chosen as is convenient.%

\subsection{Relational observables}
\label{sec:LQCrelnl}

Class operators constructed according to Eq.\ (\ref{eq:classopdefqc}) express
questions concerning the correlation between the values of the quantities
$A^{\alpha_i}$ at the specified values $\phi_i$ of the scalar field.
Probabilities computed in this way therefore naturally correspond to
predictions concerning the corresponding relational observables, such as for
the volume Dirac observable $\hat \nu|_\phi$ defined in terms of the 
propagator, Eq.\ (\ref{eq:propdefqm}), by
\begin{equation}
\hat \nu|_{\phi^*} (\phi) = U(\phi^*-\phi)^\dagger  \hat \nu \, U(\phi^* - \phi) ~.
\end{equation}
(See, for example, Eq.\ (\ref{eq:probvol}).)  In other words, as discussed in
detail in Ref.\ \cite{CS10c}, probabilities for histories of values of an
observable $\hat{A}$, which does not commute with the constraint, are
naturally expressed in terms of the corresponding relational Dirac observable
$\hat{A}|_{\phi^*}(\phi)$, which does.  In this sense the notion of relational
observables arises naturally, indeed, almost inevitably, in the framework of
consistent histories when predictions concerning correlations between values
of observables are concerned.

\section{Applications}
\label{sec:apps}

We now apply the generalized ``consistent histories'' quantum theory of
loop-quantized cosmology we have constructed to a series of predictions
concerning histories of physical quantities of interest.  In each case the
approach is the same.  Within the decoherent histories framework for quantum
prediction, for any physical question that is to be investigated the
corresponding class operators and branch wave functions must be constructed.
If the interference between these branch wave functions disappears -- i.e.\ if
the family of branch wave functions corresponding to the question decoheres --
then quantum probabilities may be assigned to the alternatives according to
the diagonal elements of the decoherence functional, the norms of the
corresponding branch wave functions.

It should be noted that in general there are a number of distinct reasons
decoherence might occur.  Predictions concerning physical quantities at a
single value of the scalar field (moment of `time') \emph{always} decohere,
because the corresponding family of class operators are simply orthogonal
projections.  (Compare for example Eq.\ (\ref{eq:decohvol}).  This is
essentially the reason the need for decoherence is not so evident in simple
applications of quantum mechanics that do not concern predictions for
\emph{sequences} of quantum events.)  More generally, decoherence might obtain
because of symmetries or selection rules; because of individual properties of
the histories in question; or because of properties of the particular quantum
state.  
In the applications we consider, we  encounter examples in which
decoherence occurs for each of these reasons.

Predictions concerning histories of values of the scalar momentum $p_{\phi}$
in sLQC -- making precise the sense in which it is a conserved quantity in the
quantum theory -- follow precisely the same pattern as in the Wheeler-DeWitt
theory, for which see Ref.\ \cite{CS10c}.
Decoherence in this case is essentially a consequence of the fact that 
the scalar momentum commutes with the constraint.

Our primary goal in this paper, however, will be to demonstrate how
probabilities for the quantum bounce of the volume observable can be computed.
We begin with the construction of class operators for the volume of the
fiducial cell of the universe at an instant of `time' $\phi$, and also a
sequence of values of $\phi$.  Predictions concerning \emph{histories} of
values of the volume are interesting because in this case decoherence is no
longer trivial, and indeed will frequently not obtain.  We will nonetheless
exhibit several physical examples in which it does, and use these to study two
important physical problems: quasiclassical behavior of the universe; and the
quantum ``bounce''.

\subsection{Class operators for volume}
\label{sec:volphi}

We begin with the class operator  for the history in which the volume $\nu$ is
in $\Delta\nu$ when the scalar field has value $\phi^*$. It is simply given by
\begin{equation}
C_{\Delta\nu|_{\phi^*}}  =
  U^{\dagger}(\phi^*-\phi_o) \Projsupb{\nu}{\Delta \nu} U(\phi^*-\phi_o),
\label{eq:classopvol}
\end{equation}
where the projection $\Projsupb{\nu}{\Delta \nu}$ is
\begin{equation}\label{eq:projvol}
P^\nu_{\Delta \nu} = \sum_{\nu \in \Delta \nu} \, |\nu\rangle \langle \nu| ~.
\end{equation}
  Note
that we employ here projections onto ranges of values of the volume operator
$\hat{\nu}$, not the Dirac observable $\hat{\nu}|_{\phi^*}$.
These ranges form a collection of disjoint sets that cover the full range of
discrete volume eigenvalues, $0 \leq |\nu| = 4 \lambda n < \infty$, such that
$\sum_i C_{{\Delta \nu_i}|\phi^*} = \Id$.

If $|\Psi\rangle$ denotes a quantum state of the universe at $\phi = \phi_o$,
the branch wave functions for these histories are
\begin{equation}
\ket{\Psi_{\Delta\v|_{\phi^*}}(\phi)} =
  U(\phi-\phi_o) C_{\Delta\v|_{\phi^*}}^{\dagger} \ket{\Psi}.
\label{eq:bwfvol}
\end{equation}
Because in this instance the class operators are simply projections, the 
branch wave functions for these histories are orthogonal,
\begin{subequations}
\begin{eqnarray}
\bracket{\Psi_{\Delta\v_i|_{\phi^*}}}{\Psi_{\Delta\v_j|_{\phi^*}}}
 &=&
\bracket{C_{\Delta\v_i|_{\phi^*}}^{\dagger} \Psi}{C_{\Delta\v_j|_{\phi^*}}^{\dagger} \Psi}
\label{eq:decohvol-a}\\
& = &
\bracket{\Psi_{\Delta\v_i|_{\phi^*}}}{\Psi_{\Delta\v_i|_{\phi^*}}}\,\cdot\, 
\delta_{ij} ~.
\label{eq:decohvol-c}
\end{eqnarray}
\label{eq:decohvol}%
\end{subequations}%
This implies that for this family of histories decoherence is automatic.  One
can thus meaningfully compute the quantum probabilities.  Using Eqs.\
(\ref{eq:classopvol}) and (\ref{eq:projvol}), the probability that the
universe has volume in the range $\Delta \nu$ when $\phi = \phi^*$ is then 
given by
\begin{subequations}
\begin{eqnarray}
p_{\Delta\v}(\phi^*) & = & 
\bracket{\Psi_{\Delta\v|_{\phi^*}}}{\Psi_{\Delta\v|_{\phi^*}}}
\label{eq:probvol-a}\\
& = & 
\melt{\Psi}{C_{\Delta\v|_{\phi^*}}^{\dagger}}{\Psi}
\label{eq:probvol-b}\\
& = & 
\melt{\Psi(\phi^*)}{\Projsupb{\v}{\Delta \v}}{\Psi(\phi^*)}
\label{eq:probvol-e}\\
& = & 
\sum_{\v\in\Delta\v} |\Psi(\v,\phi^*)|^2~.
\label{eq:probvol-f}
\end{eqnarray}
\label{eq:probvol}%
\end{subequations}

\begin{figure}[htbp!]
\subfloat[{$\Delta\v^*=[0,40\lambda]$}]{%
\includegraphics[width=0.50\textwidth]{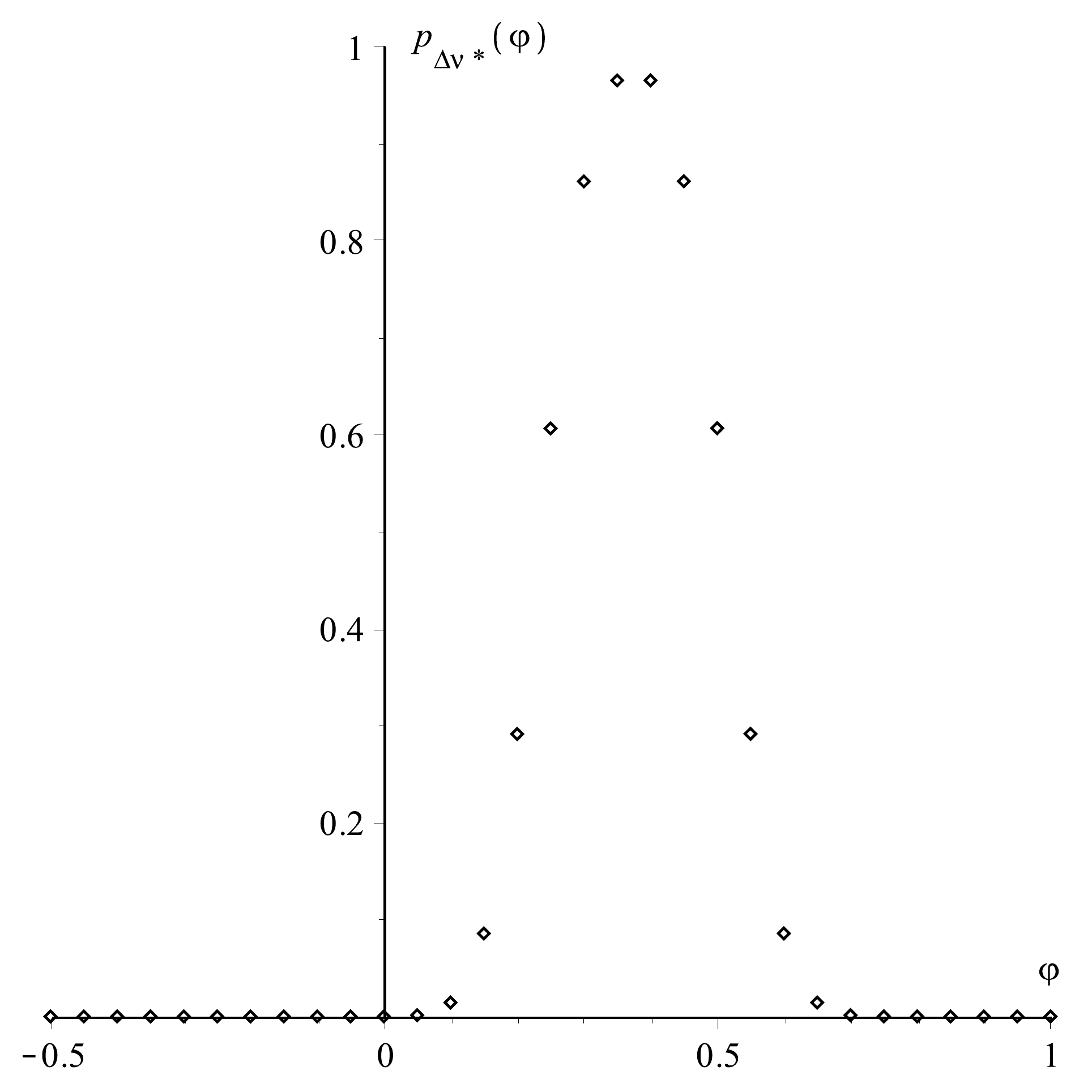}
\label{fig:probvol-a}
}%
\subfloat[{$\Delta\v=[80\lambda,120\lambda]$}]{%
\includegraphics[width=0.50\textwidth]{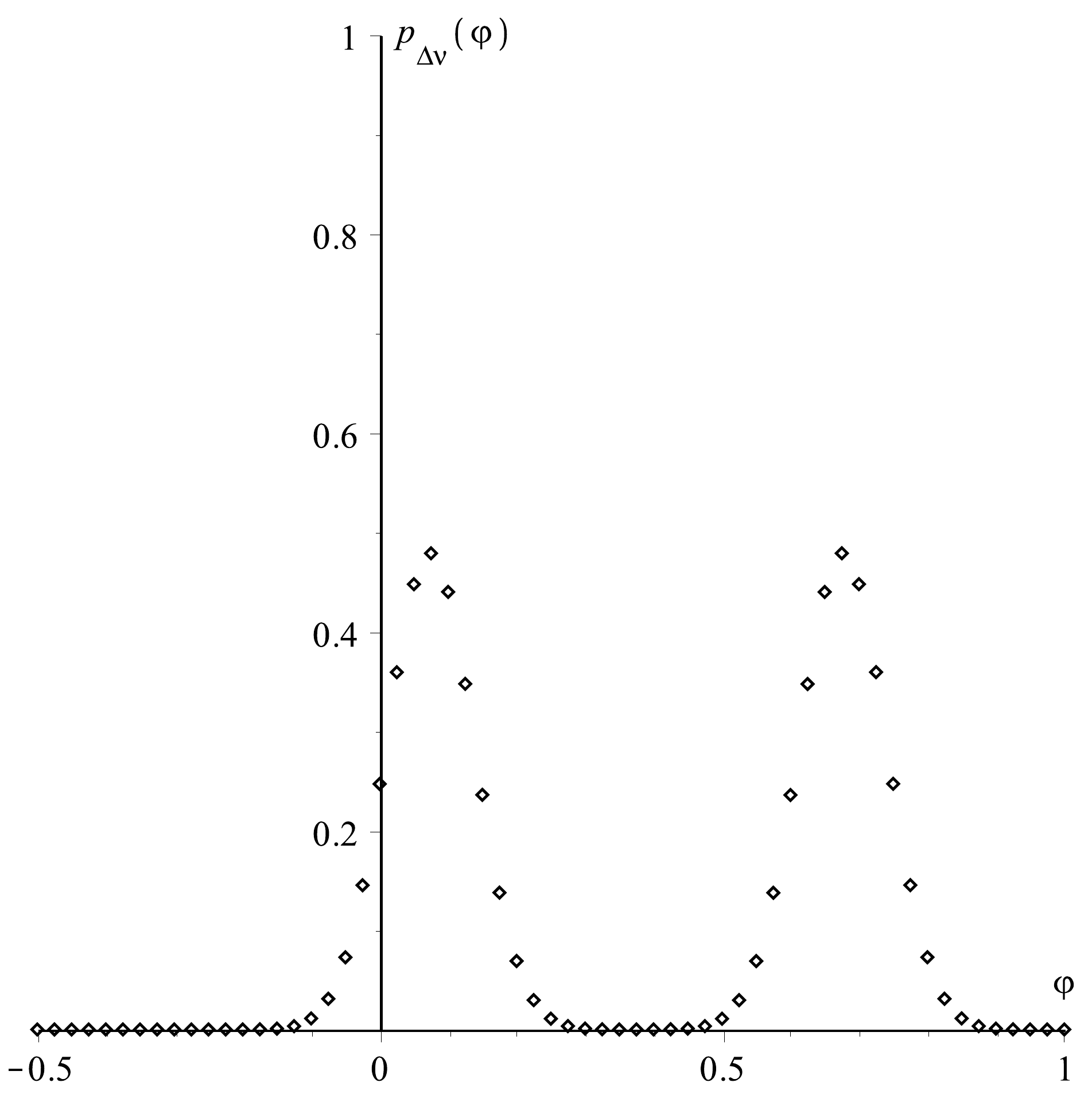}
\label{fig:probvol-b}%
}%
\caption{Plots of the probability $p_{\Delta\v}(\phi)$ that the volume of (a
fiducial cell of) a quantum universe specified by a state which is
semiclassical at large $|\phi|$ is in the range $\Delta\v$ when the scalar
field has value $\phi$ for two choices of range $\Delta\v$.  The bounce
connecting the semiclassical phases is clearly visible in both plots.  The
first plot shows the probability that the volume of the universe is less than
$\v^*=40\lambda$ as a function of $\phi$.  For this particular state the
volume at which the universe ``bounces'' is smaller than $\v^*$, and the
probability the volume of the universe is less than $\v^*$ at the bounce is
therefore close to unity, becoming zero as $|\phi|$ becomes large.  The second
plot shows the probability that the volume of the fiducial cell is in the
range $\Delta\v=[80\lambda,120\lambda]$, a range the state passes through on
both sides of the bounce.  The state shown has
$\tilde{\Psi}_{sc}(k)=1/\sqrt{2\sigma\sqrt{\pi}}
\exp(-(k-\bar{k})^2/2\sigma^2)\exp(-ik\ln|\bar{\v}/\lambda|) +
(k\leftrightarrow-k)$
in Eq.\ (\ref{eq:a_Psiexpnlim}), with $\phi_o=0$, $\bar{\v}=10\lambda$,
$\bar{k}=15$, and $\sigma=2$.  This state is peaked on a solution to the
``effective'' cosmological equations of LQC
\cite{willis04,taveras08a,ashsingh11} that approaches classical
expanding/collapsing solutions at large volume, joined by a bounce at small
volume in between -- as depicted, for example, in Figs.\
\ref{fig:LQCsc}-\ref{fig:smallvol}.  
(In those figures, however, the effective
trajectory shown happens to be symmetric about $\phi=0$, which is not the 
generic case.)
For the given parameters this state
``bounces'' at $(\v_B,\phi_B)=(30\lambda,0.375)$ in units in which $G=1$.
(For details of the correspondence between the trajectories of semiclassical
states in LQC and solutions to the effective equations see Ref.\
\cite{dac13c}.)
}%
\label{fig:probvol}%
\end{figure}

By way of example, Fig.\ \ref{fig:probvol} shows probabilities calculated from
Eq.\ (\ref{eq:probvol}) that a state which is quasiclassical at large volume%
\footnote{That is, a state for which $\Psik(k)$ in Eq.\
(\ref{eq:a_Psiexpnlim}) is a symmetric superposition of Gaussians in $k$ --
one each corresponding to the collapsing and expanding Wheeler-DeWitt branches
of the bouncing solution at large volume; see Ref.\ \cite{dac13c} for details.
} %
takes on volumes in the range $\Delta\v$ for two choices of that range.  For
example, Fig.\ \ref{fig:probvol-a} shows the probability that the universe
takes on small volume.  Specifically, the plot shows the probability as a
function of the scalar field $\phi$ that the volume of the fiducial spatial
cell has volume less than or equal to $\v^*$ i.e.\ that $|\v|\in \Delta\v^*$,
where $\Delta\v^* = [0,\v^*]$:
\begin{equation}
p_{\Delta\v^*}(\phi) = \sum_{|\v|\in\Delta\v^*} |\Psi(\v,\phi)|^2.
\label{eq:probvolsm}
\end{equation}
The quantum bounce is clearly visible in the plot, the probability the
universe has small volume becoming zero as $|\phi|$ becomes large.

We now consider the more interesting case when a sequence of `time' instants
is involved.  In contrast to the class operator representing the volume of the
universe at an instant $\phi = \phi^*$ (Eq.(\ref{eq:classopvol})), the class
operator for the volume to take particular values in ranges $\Delta \nu_i$ at
a sequence of different instances of internal time $\{\phi_1,...,\phi_n\}$ is
not a simple projection.  It is given by
\begin{equation}
C_{\Delta\v_1|_{\phi_1};\Delta\v_2|_{\phi_2};\cdots;\Delta\v_n|_{\phi_n}}
= \Projsupb{\v}{\Delta \v_1}(\phi_1) \Projsupb{\v}{\Delta \v_2}(\phi_2)
      \cdots \Projsupb{\v}{\Delta \v_n}(\phi_n),
\label{eq:classopvolseq}
\end{equation}
where the sets of ranges
$(\{\Delta\v_1\},\{\Delta\v_2\},\dots,\{\Delta\v_n\})$ partition the allowed
range of volumes.
As remarked earlier, in general it is neither obvious nor trivial that the
corresponding branch wave functions (Eq.\ (\ref{eq:bwfdefqc})) decohere.
Nevertheless, in the following we will exhibit several important (and typical)
examples for which they do, and extract the corresponding quantum
probabilities.

\subsection{Decoherence for semiclassical states}
\label{sec:qclassevol}

We first apply this framework to states which are semi-classical at late times
in this loop quantized model.  Analysis of such states in loop quantum
cosmology using sophisticated numerical simulations was first performed in
Refs.\ \cite{aps_letter,aps,aps:improved} for the spatially flat homogeneous
and isotropic model sourced with a massless scalar field.%
\footnote{Subsequently, these studies have been extended to different models
in loop quantum cosmology.  For a review of numerical studies of these models,
see Ref.\ \cite{singh12a}.
} %
The states are chosen such that they are initially peaked on classical
trajectories in a macroscopic universe, and evolved using the quantum
gravitational Hamiltonian constraint (see Eq.\ (\ref{eq:theta})).  Numerical
simulations show that such states remain peaked on classical trajectories
until the spacetime curvature reaches almost a percent of its value at the
Planck scale.  As the Planck scale is approached, significant departures arise
between the classical trajectory, Eq.\ (\ref{eq:classtraj}), and the
trajectory obtained from the expectation value of the volume observable.
Instead of reaching the classical big bang singularity, such states bounce
when the energy density of the universe reaches a maximum value
$\rho_{\rm{max}} \approx 0.41 \rho_{\rm{Planck}}$.  After the bounce, states
are found to be peaked on an expanding classical solution (disjoint in the
classical theory from the one where the initial state was peaked)
\cite{aps,aps:improved}.  This result, initially obtained using numerical
simulations for a class of semiclassical states, can be generalized to all the
states in the physical Hilbert space.  It turns out that in sLQC
the expectation value of the volume observable has a minimum irrespective of
the choice of state \cite{acs:slqc}.  Further, \emph{all} states in the
physical Hilbert space reach arbitrarily large volume in the infinite past and
future ($\phi\rightarrow\pm\infty$).  The minimum of the expectation value of
volume translates to an upper bound on the expectation values of the energy
density observable \cite{acs:slqc}.  Recently, this result has also been
understood from an analytical study of the properties of the eigenfunctions of
the gravitational constraint,
Eq.\ (\ref{eq:theta}) \cite{dac13a}.  An important result of various numerical
investigations in loop quantum cosmology is that states which are
semi-classical at late times follow an effective trajectory throughout their
evolution \cite{ashsingh11}.  The effective trajectory is derived using an
effective Hamiltonian constraint obtained via geometrical methods of quantum
mechanics \cite{willis04,taveras08a}.  As with the expectation value of the
volume observable, significant departures exist between the effective and the
classical trajectories at the Planck scale, whereas at spacetime curvatures
much smaller than the Planck value, the effective and classical trajectories
coincide.

How is the question of whether or not a given state follows a classical or an
effective trajectory posed within the framework of generalized quantum theory?
A state ``follows a trajectory'' in minisuperspace when it exhibits a
correlation between $\phi$ and $\v$ given by that trajectory with a high
probability.  The fidelity of this correlation may be specified with varying
degrees of precision.  To accomplish this, we consider a coarse-graining of
minisuperspace on a set of slices $\{\phi_1,\phi_2,\cdots,\phi_n\}$ by 
positive ranges of volume $\{\Delta\v_{i_k},k=1\ldots n\}$ on each slice
$\phi_k$, so that $\cup_{i_k}\Delta\v_{i_k}=[0,\infty)$ on each slice $k$.
To track a particular minisuperspace trajectory $\gamma$, choose the
partitions $\{\Delta\v_{i_k}\}$ such that one range $\Delta\v_{\gamma_k}$ from
each partition encloses $\gamma$ at each $\phi_k$.  To the degree of precision
specified by this coarse-graining, a state $\ket{\Psi}$ may be said to
``follow'' $\gamma$ with near certainty if the only branch wave function that
is not essentially zero is
\begin{subequations}
\begin{eqnarray}
\ket{\Psi_{\gamma}} & = &
U(\phi-\phi_o)\Projsupb{\v}{\Delta\v_{\gamma_n}}(\phi_n)
 \cdots\Projsupb{\v}{\Delta\v_{\gamma_2}}(\phi_2)
   \Projsupb{\v}{\Delta\v_{\gamma_1}}(\phi_1)
  \ket{\Psi}
\label{eq:Psigamma-a}\\
 & \equiv & U(\phi-\phi_o)C_{\gamma}\ket{\Psi}.
\label{eq:Psigamma-b}
\end{eqnarray}
\label{eq:Psigamma}%
\end{subequations}
If indeed the branch wave function for the complementary history
$\bar{\gamma}$ (``does not follow $\gamma$'') vanishes,
\begin{equation}
\ket{\Psi_{\bar{\gamma}}}  \equiv  U(\phi-\phi_o)(\Id-C_{\gamma})\ket{\Psi}
  \approx  0,
\end{equation}
\label{eq:Psigammabar}%
then the partition $(\gamma,\bar{\gamma})$ -- i.e.\ (``follows
$\gamma$'',``does not follow $\gamma$'') -- decoheres, and $\ket{\Psi}$ may be
said to follow the trajectory $\gamma$ with probability 1.  Put another way,
the state $\ket{\Psi}$ may be said to exhibit the pattern of correlation
between volume and scalar field specified by the trajectory $\gamma$ with a high
probability.

\begin{figure}[tbh!]
\includegraphics[width=0.75\textwidth]{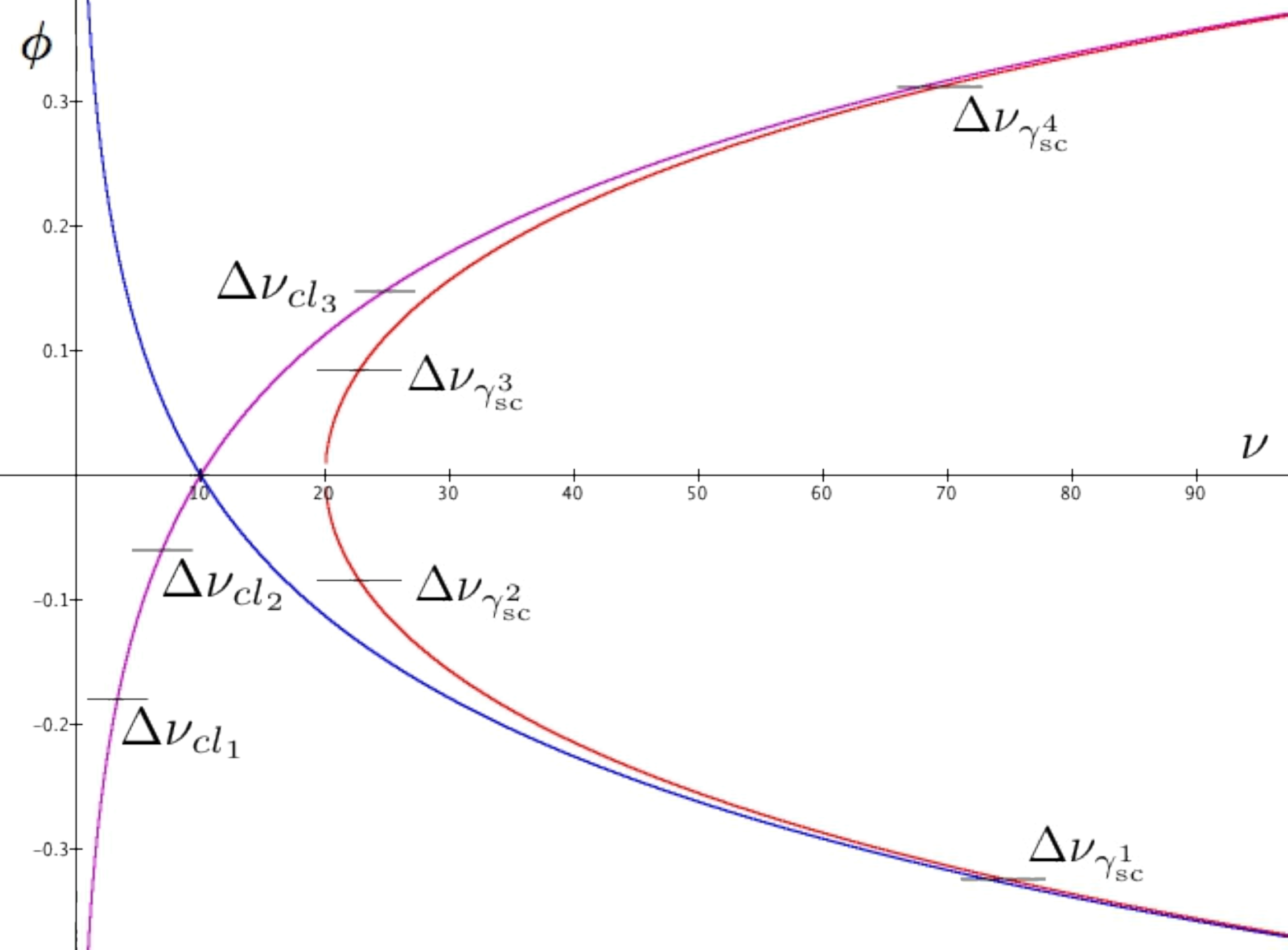}
\caption{This plot depicts coarse-graining by ranges of values of the volume
at different values of the scalar field for two histories.  The first is a
coarse-grained history
$(\Delta\v_{{cl}_1},\Delta\v_{{cl}_2},\Delta\v_{{cl}_3})$ describing an
expanding universe peaked on an expanding classical trajectory.  The second
history
$(\Delta\v_{\gamma_{\text{sc}}^1},\Delta\v_{\gamma_{\text{sc}}^2},\ldots)$
describes a 
trajectory in loop quantum cosmology, characterized by a symmetric bounce,
which is peaked on symmetrically related expanding and collapsing classical
trajectories at large $|\phi|$.  Such bouncing trajectories for states which
are semiclassical at large $|\phi|$ can be obtained from the effective
Hamiltonian approach in LQC \cite{ashsingh11}, which leads to modified 
versions of the standard Friedmann and Raychaudhuri equations.  }
\label{fig:LQCsc}
\end{figure}

Even for a state \emph{centered} on $\gamma$, whether or not
$\ket{\Psi_{\bar{\gamma}}}\approx 0$ will depend on the width of the intervals
$\Delta\v_{\gamma_k}$ relative to the width of the state $\Psi(\v,\phi_k)$ at
each $\phi_k$.  Trying to specify the path too narrowly will lead to a
partition which fails to decohere and must be further coarse-grained (by
combining some of the intervals surrounding the $\Delta\v_{\gamma_k}$) to
regain decoherence, and therefore the means to define probabilities
consistently.  Thus, as is usual in quantum theory, attempting to specify a
path \emph{too} precisely leads to a loss of predictability.  For further
discussion, see Ref.\ \cite{CS10c}.

In loop quantum cosmology, as noted above, numerical simulations show that
states $\ket{\Psi_{\text{sc}}}$ which are semi-classical%
\footnote{There are various ways to define such states.  A simple way is to
consider a Gaussian state at large volume constructed using the eigenfunctions
of the gravitational constraint operator
$\hat\Theta^{{\scriptscriptstyle\text{WdW}}}$ in the Wheeler-DeWitt theory
analogous to Eq.\ (\ref{eq:theta}).  Another way is to compute the
eigenfunctions of the $\hat \Theta$ operator either numerically (as done eg.\
in Refs.\ \cite{aps,aps:improved}) or analytically \cite{dac13a,dac13c}, and
use them to construct a semi-classical state.
} %
at early times on a contracting branch
are peaked on classical solutions at large volume and connect to the expanding
branch smoothly through a ``bounce'' in the Planck regime.  Such states are
peaked on a trajectory which is a solution to the modified Friedmann and
Raychaudhuri equations of LQC noted above for the entire evolution.  If
$\gamma_{\text{sc}}$ is chosen to be such an effective trajectory in Fig.\
\ref{fig:LQCsc}, and the widths $\Delta\v_{\gamma_k}$ chosen to be wider than
the width of $\Psi_{\text{sc}}(\v,\phi_k)$ at each $\phi_k$,%
\footnote{The dispersions of states in sLQC are discussed in detail in Refs.\
\cite{cor-singh08a,kp10a,cormont11a}.
} %
then essentially the only non-zero branch wave function will be the state
$\ket{\Psi_{\gamma_{\text{sc}}}}$ of Eq.\ (\ref{eq:Psigamma}), and
$\ket{\Psi_{\text{sc}}}$ follows the trajectory $\gamma_{\text{sc}}$ with
probability 1.

The origin of this behavior can also be analytically understood via the
dynamical eigenstates $e^{(s)}_k(\v)$ \cite{dac13a}.  \emph{All} states in
sLQC -- whether semiclassical or not -- approach a particular symmetric
superposition of expanding and contracting Wheeler-DeWitt universes at large
volume.
If the state is chosen in such a way that it is peaked on a collapsing
classical trajectory at large volume as $\phi\rightarrow-\infty$ (say), then
this state will be peaked on a corresponding expanding classical trajectory as
$\phi\rightarrow+\infty$.  (See Ref.\ \cite{dac13c} for further details.)  The
asymptotic behavior of the eigenfunctions dictates the symmetric nature of the
bounce.

\subsection{Singularity avoidance in loop quantum cosmology}
\label{sec:bounce}

We have already discussed the manner in which semi-classical states which are
peaked on classical trajectories in a large macroscopic universe at early
times bounce at a finite volume in LQC, connecting collapsing and expanding
classical solutions.  In this way, such states avoid the classical singularity
at zero volume.  As first shown analytically in Ref.\ \cite{acs:slqc}, this
behavior is generic: the expectation value of the volume is bounded below
for \emph{all} states (in the domain of the physical operators) in sLQC.

In this subsection we discuss the quantum bounce for generic states from the
perspective of consistent histories.
In Ref.\ \cite{CS10c}, the problem of the singularity in a Wheeler-DeWitt
quantized flat scalar Friedmann-Lema\^{i}tre-Robertson-Walker cosmology was
addressed through a study of the volume observable.  There it was shown that
for \emph{any} choice of fixed volume $V^*$ of the fiducial cell, the volume
of the quantum universe would invariably fall below it with unit probability.
The Wheeler-DeWitt universes are therefore inevitably singular in the sense
that they assume arbitrarily
small volume at some point in their history.%
\footnote{Ref.\ \cite{acs:slqc} showed previously that the expectation value
of the volume observable vanishes when $\phi \rightarrow -\infty$ in an
expanding branch, and the matter density always diverges in these 
Wheeler-DeWitt quantized models.
} %

In this analysis, the role of a proper understanding of quantum \emph{history}
proved crucial.  As noted, the loop quantization of this model yields states 
which are a symmetric superpositions of expanding and contracting
cosmologies at large volume.  Ref.\ \cite{CS10c} therefore analyzed a
superposition of expanding and contracting Wheeler-DeWitt universes with an
eye toward the question of whether this superposition itself could in some
sense be the \emph{reason} for the bounce.  Calculation of the probability
that the universe is found at small volume for such a superposition reveals
that at \textbf{any} given value $\phi$ of the scalar field, the probability
that the universe has volume less than $V^*|_{\phi}$ is in general between 0
and 1.  The probability that the universe is \emph{not} at arbitrarily small
volume at $\phi=-\infty$ or $\phi=+\infty$ is therefore in general not 0.
Naively this suggests the possibility that a superposition of expanding and
contracting Wheeler-DeWitt universes has a non-zero probability of being at
non-zero volume at \emph{both} $\phi=-\infty$ and $\phi=+\infty$ i.e.\ that
there is a non-zero probability of a quantum bounce.

A more careful consistent histories analysis showed that this naive
possibility is not realized.  The physical statement that the universe
``bounces'' is the statement that the volume of the universe is large at
\emph{both} $\phi=-\infty$ and $\phi=+\infty$.  A proper characterization of
the bounce is therefore a statement about the volume of the universe at a
\emph{sequence} of values of $\phi$ -- a history.  Ref.\ \cite{CS10c} shows
that for generic initial states the histories corresponding to the
alternatives $\{\text{bounce},\text{singular}\}$ decohere in the limit
$|\phi|\rightarrow\infty$ %
so that probabilities may be consistently assigned to them, and that the
probability for the bouncing history $p_{\text{bounce}}=0$: even
superpositions of expanding and contracting Wheeler-DeWitt universe cannot
bounce for any choice of state.%
\footnote{See also Ref.\ \cite{halliwell89} for early work related to this
question.
} %

We now show in detail that, in sharp contrast to the case of the
Wheeler-DeWitt quantization, the probability that \emph{generic} quantum
states in sLQC are at small volume as $\phi\rightarrow\pm\infty$ is zero.  In
fact, for any choice of volume $V^*$, we show in a sense to be made precise 
below that the probability the volume
of the universe is larger than $V^*$ is unity as $|\phi|\rightarrow\infty$: 
\emph{all} states in this model achieve arbitrarily large
volume in both limits.  In this sense every state retains some flavor of
the striking ``bounce'' of the narrowly peaked quasi-classical ones.

Next, we address histories of the volume with evolution in $\phi$.  We show
that for arbitrary quantum states the family of coarse-grained alternative
histories $\{\text{bounce},\text{singular}\}$ decoheres, as in the
Wheeler-DeWitt case.  However, in contrast to the Wheeler-DeWitt case, the
probability that the universe is singular in the scalar past or future is
zero, and the probability that the universe bounces, unity.  \emph{All}
states in sLQC bounce from arbitrarily large volume in the ``past''
($\phi\rightarrow-\infty$) to arbitrarily large volume in the ``future''
($\phi\rightarrow+\infty$).

As in the Wheeler-DeWitt theory, it is worth emphasizing the role of the limit
$\phi\rightarrow\pm\infty$.  One may expect that for wide classes of states
such as localized states with certain peakedness properties decoherence obtains to a high degree
of approximation at finite $\phi$.  Nonetheless, it is only in the limit
$\phi\rightarrow\pm\infty$ that we are guaranteed decoherence, and hence a
bounce with probability 1, for \emph{all} states, and therefore -- in that
limit -- that a bounce is a universal prediction of the theory.

\vspace{10pt}
\textbf{REMARK:}  In Refs.\ \cite{fpps12,*pf13a} it is argued that the 
consistent histories approach to quantum theory is insufficient to address 
questions such as whether a quantum bounce takes place because histories 
involving `genuine' quantum states are inconsistent when more than two 
moments of (scalar) time are involved, or in other words, that in this case 
only histories for semiclassical states decohere.   We do not agree.

The basis for this argument is a nice calculation (in the Wheeler-DeWitt
quantization) along the lines of the one we perform in Ref.\ \cite{CS10c} and
below of the interference between histories characterized by the alternatives
$\{\text{bounce},\text{singular}\}$ in both the infinite scalar past and
future, but with a third projection onto these alternatives at an arbitrary
intermediate $\phi$.  The authors calculate a representative off-diagonal
(interference) matrix element of the decoherence functional and argue that it
is zero if and only if the corresponding state is semiclassical in the sense
that it is sharply peaked on a classical trajectory.  Unfortunately, 
it is easy to generate a wide range of counter-examples to the claim that 
the calculated matrix element is zero only for semiclassical states.  
Therefore, it is simply not the case that it is only for semiclassical states
that families of histories that study the bounce at more than two values of
scalar time decohere.  We do, however, expect the calculation the authors of
Refs.\ \cite{fpps12,*pf13a} give of the decoherence functional itself for such
three `time' histories to be useful.

Moreover, it is probably worth a certain emphasis that there are many
instances in which one would \emph{not} expect decoherence of a family of
histories, and indeed, would be suspicious of a quantum theory that purports
to do so.  Far from being a defect of the theory, it is a necessary
requirement of a theory that reproduces the predictions (or absence thereof)
of quantum theory without the introduction of e.g.\ non-local hidden
variables.  (For the purposes of this remark we include the de Broglie-Bohm
formulation in this class of theories.)  The two-slit experiment is the
classic example: any theory which assigns observationally verifiable
probabilities to the individual paths the electron follows when the physical
setup is not such that which-path information is gathered is not quantum
mechanics, and indeed will have a difficult time reproducing the predictions
of quantum mechanics absent such non-local modifications.  However, when there
are additional degrees of freedom (such as a gas of air molecules in the
two-slit apparatus) which might carry a record of which-path information,
decoherence is to be expected and probabilities for individual paths may be
assigned.  In a similar way, in cosmological models with realistic
inhomogeneous matter degrees of freedom (for example), one would expect
decoherence of histories for bulk variables like the volume for most quantum
states.

\subsubsection{Probability for zero volume in sLQC}

\begin{figure}[tbh!]
\includegraphics[width=0.75\textwidth]{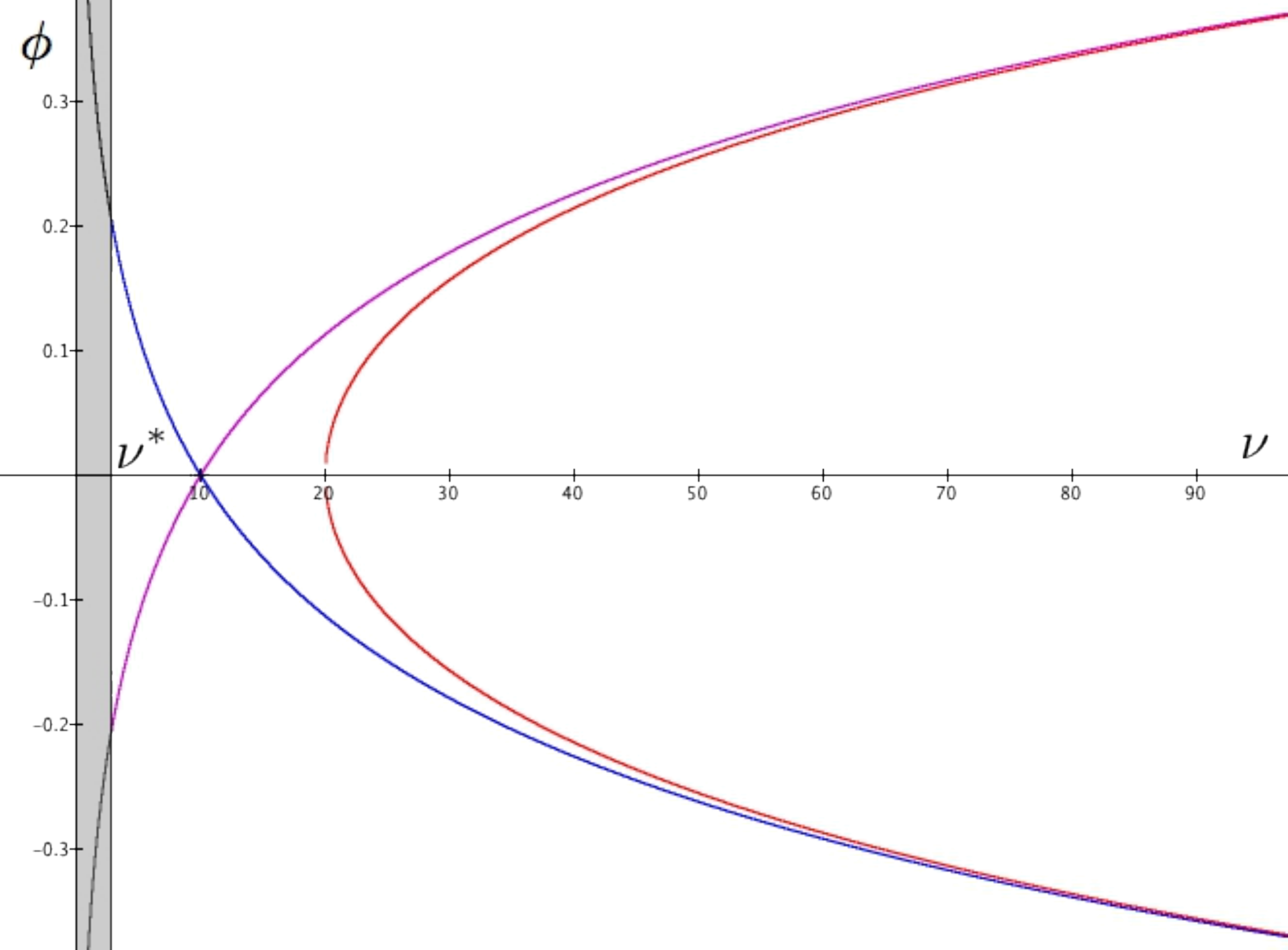}
\caption{Coarse-graining of minisuperspace suitable for studying the
probability that the universe assumes large or small volume.  Partition the
volume $\v$ into the range $\Delta\v^*=[0,\v^*]$ (the shaded region in the
figure) 
and its complement $\overline{\Delta\v^*}=(0,\infty)$.  The quantum universe
may be said to attain small volume if the probability for the branch wave
function $\ket{\Psi_{\smash{\Delta\v^*}}(\phi)}$ is near unity while that for
$\ket{\Psi_{\smash{\overline{\Delta\v^*}}}(\phi)}$ is near zero for arbitrary
choices of $\nu^*$.  Conversely, the universe may be said to attain
arbitrarily large volume over some range of $\phi$ if the probability for
$\ket{\Psi_{\smash{\overline{\Delta\v^*}}}(\phi)}$ is near unity for arbitrary
choice of $\v^*$ over that range of $\phi$.  Note that in sLQC, unlike in the
Wheeler-DeWitt quantization of the same model, volume is discrete.  
}
\label{fig:smallvol}
\end{figure}

Following Ref.\ \cite{CS10c}, one way to approach the question of whether a
quantum universe is in some sense singular is to ask whether it achieves zero
volume at any point in its evolution.%
\footnote{Since the matter density is, classically, the ratio of the (square
of the) scalar momentum -- a constant of the motion -- and the (square of the)
volume, one expects that zero volume corresponds to a diverging matter density
in the quantum theory.  This expectation is born out by the structure of the
proofs of Refs.\ \cite{acs:slqc,dac13a} that the matter density is bounded
above in sLQC.
} %
In Sec.\ \ref{sec:volphi} we showed how to calculate the probability that the
volume falls in a range specified by $\Delta\v=[\v_1,\v_2]$.  To ask whether
the volume of the universe is ever small we choose a reference volume $\v^*$
and partition the volume into the range $\Delta\v^*=[0,\v^*]$
and its complement,
$\overline{\Delta\v^*}=(\v^*,\infty)$.
The universe then has small volume at scalar time $\phi^*$ if
$|\v|\in\Delta\v^*$ at $\phi^*$ and not if $|\v|\in\overline{\Delta\v^*}$.
(See Fig.\ \ref{fig:smallvol}.)  The class operators for these alternatives
are simply the (Heisenberg) projections given by Eq.\ (\ref{eq:classopvol})
with $\Delta\v=\Delta\v^*,\overline{\Delta\v^*}$ and corresponding branch wave
functions, Eq.  (\ref{eq:bwfvol}).  The probabilities are given by Eq.\
(\ref{eq:probvol}).  Thus
\begin{subequations}
\begin{eqnarray}
\ket{\Psi_{\Delta\v^*|_{\phi^*}}(\phi)} & = &
  U(\phi-\phi^*)\Projsupb{\v}{\Delta\v^*}\ket{\Psi(\phi^*)}
\label{eq:bwfvols-a}\\
 & = & U(\phi-\phi^*)\sum_{|\v|\in\Delta\v^*}\ket{\v}\, \Psi(\v,\phi^*)
\label{eq:bwfvols-b}
\end{eqnarray}
\label{eq:bwfvols}%
\end{subequations}
and
\begin{equation}
p_{\Delta\v^*}(\phi) = \sum_{|\v|\in\Delta\v^*}|\Psi(\v,\phi)|^2,
\label{eq:probvols}
\end{equation}
where since $\phi^*$ is arbitrary we have set $\phi^*=\phi$ in the expression
for the probability.

In order to compute this probability, we will use some key properties of the
symmetric eigenfunctions of the gravitational constraint operator $\hat \Theta$ 
of Eq.\ (\ref{eq:theta}), labelled by
$k\in(-\infty,\infty)$: 
\begin{equation}
 \hat \Theta e_k^{(s)}(\nu) = \omega_k^2 e_k^{(s)}(\nu),
\end{equation} 
where $\omega_k$ is related to $p_\phi$ and $k$ by $p_\phi = \pm\hbar\omega_k$
and $\omega_k = \sqrt{12 \pi G} |k|$, respectively.  The symmetric
eigenfunctions are real and satisfy $e_{-k}^{(s)}(\nu) = e_k^{(s)}(\nu)$ and
$e_k^{(s)}(-\nu) = e_k^{(s)}(\nu)$ \cite{aps:improved,dac13a}.  A notable
property of these eigenfunctions is that they decay exponentially to zero for
volumes smaller than a cutoff value proportional to the value of $\omega_k$.
This result was first obtained in numerical simulations \cite{aps:improved},
and subsequently derived analytically in Ref.\ \cite{dac13a}, in which it was
shown that the cutoff occurs along the lines $|k| = |\nu|/2 \lambda$.
Thus, one can consider $|k| = |\nu|/2 \lambda$ as an ultra-violet momentum
space cutoff in sLQC. The exponential decay of the eigenfunctions coincides
with the volume at which the energy density attains a maximum value and the
universe bounces; the linear scaling of the cutoff with volume is what leads
to a universal maximum matter density that is independent of the quantum
state.

The symmetric eigenfunctions $e_k^{(s)}(\nu)$ satisfy
\begin{equation}
\sum_{\nu = 4 \lambda n} e_k^{(s)}(\nu) e_{k'}^{(s)}(\nu) = \delta^{(s)}(k,k')
\end{equation}
and 
\begin{equation}
\int^{+\infty}_{-\infty} \d k e_k^{(s)}(\nu) e_{k}^{(s)}(\nu') 
  = \delta_{\nu,\nu'} ~.
\end{equation}
Physical states $\Psi$ in sLQC can be constructed using the eigenfunctions
$e_k^{(s)}(\nu)$,
\begin{equation}\label{eq:a_Psiexpnlim}
\Psi(\v,\phi)  =  \int_{-\infty}^{\infty}dk\,
   \Psik(k) e^{(s)}_k(\v) e^{i\omega_k\phi},
\end{equation}
where we have set $\phi_o=0$ for convenience.  As a consequence of the
ultra-violet cutoff on the eigenfunctions, 
\begin{equation}\label{eq:Psiexpnlim}
\Psi(\v,\phi) \cong  \int_{-\v/2\lambda}^{\v/2\lambda}dk\,
\end{equation}
Note in Eq.  (\ref{eq:bwfvols}) that $\v$ is bounded by $\v^*$.  Further,
the $e^{(s)}_k(\v)$ are well-behaved functions of $k$ for all values of $\v$.
For any fixed value of $\v$, their rate of oscillation in $k$ is fixed by
$\v$.%
\footnote{In fact, this rate is approximately constant in $\v$, as can be seen
by observing that the $e^{(s)}_k(\v=4\lambda n)$ have $2n-1$ nodes in the full
range of oscillation $|k|\lesssim 2 n$ \cite{dac13a}.  These functions
therefore exhibit precisely $n$ full oscillations over that range, giving a
mean oscillation frequency $n/\Delta k \cong n/4n = 1/4$, independent of the
volume $n$, and in particular exhibit only a finite number of oscillations in
any fixed range of $k$.  The oscillations in $k$ for any given $n$ do tend to
become more rapid for smaller $|k|$, and in the regime $|\v|\gg\lambda|k|$ the
angular frequency of oscillation in $|k|$ grows as $\ln|\v/\lambda|$ i.e.\
grows slowly with $\v$.
} %

Thus, for large $|\phi|$ -- meaning at a minimum $\omega_k |\phi| \gg 1$
-- rapid oscillation of the factor $\exp(i\omega_k\phi)$ will according to the
Riemann-Lebesgue lemma eventually suppress the integral, and we find
\begin{equation}
\lim_{\substack{|\phi|\rightarrow\infty\\ \v\text{ fixed}}} \Psi(\v,\phi) = 0
\label{eq:psilim}
\end{equation}
for any fixed value of $\v \leq \v^*$.  In other words, for any fixed
$\v$, $|\phi|$ eventually becomes large enough to suppress the state, driving
the state to larger volume as $|\phi|$ increases.   Thus $\Psi(\v,\phi)$ for
$\v<\v^*$ will always be suppressed for large enough $|\phi|$ for arbitrary 
states in the theory.

On the other hand, for the complementary branch wave function
$\ket{\Psi_{\smash{\overline{\Delta\v^*}|_{\phi^*}}}(\phi)}$ corresponding to
large volume universes, $\v$ can be arbitrarily large.  As $\v$ becomes larger
a wider range of $k$'s can contribute nontrivially to the integral in Eq.\
(\ref{eq:Psiexpnlim}).  For $\v \gg 2\lambda|k|$ the rate of oscillation of
the $e^{(s)}_k(\v)$ with $k$ is fixed by the asymptotic limit of the symmetric
eigenfunctions, increasing in proportion with $\ln|\v|$ \cite{dac13a}.
Again, for any fixed $\v$ the state is suppressed as
$|\phi|\rightarrow\infty$, so that the region of support of this branch wave
function in the $(\v,\phi)$ plane must have $\ln|\v|$ increasing in proportion
with $|\phi|$ -- just the behavior of Wheeler-DeWitt quantized states.

We find, therefore, that since $\Psi(\v,\phi)$ at any fixed $\v$ vanishes in
the limit $|\phi|\rightarrow\infty$,
\begin{equation}
\lim_{\phi^*\rightarrow-\infty}\ket{\Psi_{\Delta\v^*|_{\phi^*}}(\phi)} =0
\qquad\text{and}\qquad
\lim_{\phi^*\rightarrow+\infty}\ket{\Psi_{\Delta\v^*|_{\phi^*}}(\phi)} =0.
\label{eq:bwfvolslim}
\end{equation}
As the intervals $\Delta\v^*$ and $\overline{\Delta\v^*}$ are complementary,
this implies
\begin{equation}
\lim_{\phi^*\rightarrow-\infty}\ket{\Psi_{\smash{\overline{\Delta\v^*}|_{\phi^*}}}(\phi)} =
     \ket{\Psi(\phi)}
\qquad\text{and}\qquad
\lim_{\phi^*\rightarrow+\infty}\ket{\Psi_{\smash{\overline{\Delta\v^*}|_{\phi^*}}}(\phi)} =
     \ket{\Psi(\phi)}.
\label{eq:bwfvolsbarlim}
\end{equation}
As a consequence, one finds for the probabilities
\begin{subequations}
\begin{align}
\lim_{\phi\rightarrow-\infty} p_{\Delta\nu^*}(\phi) &= 0
&
\lim_{\phi\rightarrow+\infty} p_{\Delta\nu^*}(\phi) &= 0
\label{eq:probvolslim-a}\\
\lim_{\phi\rightarrow-\infty} p_{\overline{\Delta\nu^*}}(\phi) &= 1
&
\lim_{\phi\rightarrow+\infty} p_{\overline{\Delta\nu^*}}(\phi) &= 1.
\label{eq:probvolslim-b}
\end{align}
\label{eq:probvolslim}%
\end{subequations}
We can see already from this that loop quantum states invariably bounce:
the probability the universe is found at small volume as
$|\phi|\rightarrow\infty$ is zero, regardless of the state.

Eqs.\ (\ref{eq:probvolslim}) say that \emph{all} states in sLQC achieve
arbitrarily large volume in each of the limits $\phi\rightarrow-\infty$,
$\phi\rightarrow+\infty$.  States in sLQC don't merely refrain from becoming
singular.  They inevitably grow to large volume, no matter how non-classical
the state.  (This result complements that of Ref.\ \cite{acs:slqc} that the
expectation value of the volume becomes infinite in those limits for all
states.)  In the next section we will use this to show that the family of
histories describing a quantum bounce decoheres, and that indeed all states in
the theory bounce \emph{from} arbitrarily large volume \emph{to} arbitrarily
large volume.

Finally, we observe that as the state $\ket{\Psi}$ was arbitrary and
\begin{equation}
\Projsupb{\v}{\Delta\v^*}(\phi) + \Projsupb{\v}{\overline{\Delta\v^*}}(\phi) = \Id,
\label{eq:pdeltavscomplete}
\end{equation}
Eqs.\ (\ref{eq:bwfvolslim}-\ref{eq:bwfvolsbarlim}) may be conveniently
expressed in terms of volume projections as
\begin{subequations}
\begin{align}
\lim_{\phi\rightarrow-\infty} \Projsupb{\v}{\Delta\v^*}(\phi)  &= 0
&
\lim_{\phi\rightarrow+\infty} \Projsupb{\v}{\Delta\v^*}(\phi)  &= 0
\label{eq:projvollim-a}\\
\lim_{\phi\rightarrow-\infty} \Projsupb{\v}{\overline{\Delta\v^*}}(\phi) &= \Id
&
\lim_{\phi\rightarrow+\infty} \Projsupb{\v}{\overline{\Delta\v^*}}(\phi) &= \Id
\label{eq:projvollim-b}
\end{align}
\label{eq:projvollim}%
\end{subequations}
on all states in the theory.%
\footnote{Indeed, one may carry out the argument made above in terms of
$\Psi(\v,\phi)$ directly at the level of the Heisenberg projections by
expanding the propagators in the projections in symmetric eigenkets
$\ket{k^{(s)}}$ and using $e_k^{(s)}(\nu) = \langle \nu|k^{(s)}\rangle$.  One
arrives in just the same way at Eqs.\ (\ref{eq:projvollim}), but the
formul\ae\ getting there are somewhat messier simply because one is dealing
with operators rather than states.
\label{foot:Projvlim}
} %

Note we have not so far addressed the question of whether the universe
\emph{ever} assumes volumes in $\Delta\v^*$ with non-zero probability.
In fact, examination of Eq.\ (\ref{eq:probvols}) should be sufficient to show
that so long as $\v^*>0$, there always exist states for which it will.  (See,
for example, Fig.\ \ref{fig:probvol}.)  However, this is \emph{not} sufficient
to show that the universe might become singular in sLQC. Recall that the
eigenfunctions $e^{(s)}_k(\v)$ decay exponentially for volumes smaller than
$|\nu| = 2 \lambda |k|$ and vanish at $\nu = 0$, and thus, from Eq.\
(\ref{eq:probvols}) the probability that \emph{any} state in sLQC assumes
precisely zero volume is \textbf{zero},
\begin{equation}
p_{\v=0}(\phi) =0.
\label{eq:probvol0}
\end{equation}
This result stands in sharp contrast with the situation in Wheeler-DeWitt
theory, where the rapid oscillations in the eigenfunctions as $\v\rightarrow0$
inevitably `draw in' Wheeler-DeWitt states to zero volume and infinite
density, and the probability for a singularity turns out to be non-vanishing
-- and indeed, is unity for all states in the limits $|\phi| \rightarrow
\infty$ \cite{CS10c}.

\subsubsection{Quantum bounce}
\label{sec:qbounce}

It is tempting to conclude that Eqs.\ (\ref{eq:probvolslim}) are sufficient to
demonstrate that all states in sLQC ``bounce'' \emph{from} large volume as
$\phi\rightarrow-\infty$ \emph{to} large volume as $\phi\rightarrow+\infty$.
However, as emphasized in Refs.\ \cite{CS10a,CS10b,CS10c}, statements
concerning a quantum bounce are inherently assertions concerning the volume at
a \emph{sequence} of values of $\phi$, and, as in the two-slit experiment, it
is in precisely such situations that decoherence becomes critical in order to
arrive at consistent quantum predictions.  Indeed, in Ref.\ \cite{CS10c} it is
shown that consideration of the single-$\phi$ volume probability
$p_{\Delta\nu^*}(\phi)$ alone for Wheeler-DeWitt states which are
superpositions of expanding and contracting universes may lead one to the
incorrect conclusion that a bounce is possible in that model.  However, a
proper analysis of the histories describing a quantum bounce shows that this
naive conclusion based on single-$\phi$ probabilities is misleading, and that
indeed the probability for a bounce is zero.%

How, then, is a ``bounce'' characterized within quantum theory?  The assertion
that a universe bounces is the statement that the universe assumes large
volume at both ``early'' ($\phi\rightarrow-\infty$) and ``late''
($\phi\rightarrow+\infty$) values of $\phi$.  A (highly coarse-grained)
description of a bounce may therefore be obtained by making a choice of
$\phi$-slices $\phi_1$ and $\phi_2$ and volume partitions
$(\Delta\v_1^*,\overline{\Delta\v_1^*})$ and
$(\Delta\v_2^*,\overline{\Delta\v_2^*})$ on them.  The class operator for the
history in which the universe ``bounces'' between $\phi_1$ and $\phi_2$ --
i.e. is at large volume at both  $\phi_1$ and $\phi_2$ -- is then
\begin{equation}
C_{\mathrm{bounce}}(\phi_1,\phi_2) \,=\,
C_{\overline{\Delta\nu^*_1};\overline{\Delta\nu^*_2}} \, = \, \Projsupb{\nu}{\overline{\Delta\nu^*_1}}(\phi_1)
\Projsupb{\nu}{\overline{\Delta\nu^*_2}}(\phi_2).
\label{eq:Cbouncedef}\\
\end{equation}

On the other hand, the class operator for the alternative history that
the universe is found at small volume at either or both of $\phi_1$, $\phi_2$
is
\begin{subequations}
\begin{eqnarray}
C_{\mathrm{sing}}(\phi_1,\phi_2)  & = & \Id - C_{\mathrm{bounce}}(\phi_1,\phi_2)
\label{eq:Csingdef-a}\\
& = &
C_{\smash{\Delta\nu^*_1;\Delta\nu^*_2}} +
C_{\smash{\Delta\nu^*_1;\overline{\Delta\nu^*_2}}} +
C_{\smash{\overline{\Delta\nu^*_1};\Delta\nu^*_2} }.
\label{eq:Csingdef-c}
\end{eqnarray}
\label{eq:Csingdef}%
\end{subequations}
It is clear from Eq.\ (\ref{eq:Csingdef-c}) that
$C_{\mathrm{sing}}(\phi_1,\phi_2)$ encodes the various ways the universe can
be at small volume at $\phi_1$ and/or $\phi_2$.

We now demonstrate that the only branch wave function which is non-vanishing
in the limits $\phi_1 \rightarrow - \infty$ and $\phi_2 \rightarrow \infty$ is
the one corresponding to the bounce.

Using Eqs.\ (\ref{eq:projvollim}), one finds,
\begin{subequations}
\begin{eqnarray}
\lim_{\substack{\phi_1\rightarrow -\infty\\ \phi_2\rightarrow +\infty }}
  \Projsupb{\raisebox{2pt}{$\scriptstyle{\v}$}}{\smash{\Delta\v_2^*}}(\phi_2)
  \Projsupb{\raisebox{2pt}{$\scriptstyle{\v}$}}{\smash{\Delta\v_1^*}}(\phi_1)
     \ket{\Psi}
 & = & \lim_{\phi_2\rightarrow+\infty}
           \Projsupb{\raisebox{2pt}{$\scriptstyle{\v}$}}{\smash{\Delta\v_2^*}}(\phi_2) \cdot 0
\nonumber\\
 & = & 0;
\label{eq:PPpsi-a}\\ \nonumber\\
\lim_{\substack{\phi_1\rightarrow -\infty\\ \phi_2\rightarrow +\infty }}
  \Projsupb{\raisebox{2pt}{$\scriptstyle{\v}$}}{\smash{\overline{\Delta\v_2^*}}}(\phi_2)
  \Projsupb{\raisebox{2pt}{$\scriptstyle{\v}$}}{\smash{\Delta\v_1^*}}(\phi_1) \ket{\Psi}
 & = & \lim_{\phi_2\rightarrow+\infty}
  \Projsupb{\raisebox{2pt}{$\scriptstyle{\v}$}}{\smash{\overline{\Delta\v_2^*}}}(\phi_2) \cdot 0
\nonumber\\
 & = & 0;
\label{eq:PPpsi-b}\\ \nonumber\\
\lim_{\substack{\phi_1\rightarrow -\infty\\ \phi_2\rightarrow +\infty }}
  \Projsupb{\raisebox{2pt}{$\scriptstyle{\v}$}}{\smash{\Delta\v_2^*}}(\phi_2)
  \Projsupb{\raisebox{2pt}{$\scriptstyle{\v}$}}{\smash{\overline{\Delta\v_1^*}}}(\phi_1) \ket{\Psi}
 & = & \lim_{\phi_2\rightarrow+\infty}
           \Projsupb{\raisebox{2pt}{$\scriptstyle{\v}$}}{\smash{\Delta\v_2^*}}(\phi_2) \ket{\Psi}
\nonumber\\
 & = & 0;
\label{eq:PPpsi-c}\\ \nonumber\\
\lim_{\substack{\phi_1\rightarrow -\infty\\ \phi_2\rightarrow +\infty }}
  \Projsupb{\raisebox{2pt}{$\scriptstyle{\v}$}}{\smash{\overline{\Delta\v_2^*}}}(\phi_2)
  \Projsupb{\raisebox{2pt}{$\scriptstyle{\v}$}}{\smash{\overline{\Delta\v_1^*}}}(\phi_1) \ket{\Psi}
 & = & \lim_{\phi_2\rightarrow+\infty}
          \Projsupb{\raisebox{2pt}{$\scriptstyle{\v}$}}{\smash{\overline{\Delta\v_2^*}}}(\phi_2)
             \ket{\Psi}
\nonumber\\
 & = & \ket{\Psi}.
\label{eq:PPpsi-d}
\end{eqnarray}
\label{eq:PPpsi}%
\end{subequations}
Using Eqs.\ (\ref{eq:Csingdef}) and (\ref{eq:PPpsi}), we find that the branch
wave function for an sLQC quantum universe to encounter the singularity
vanishes,
\begin{equation}
\ket{\Psi_{\text{sing}}(\phi)}  =  U(\phi-\phi_o) \lim_{\substack{\phi_1\rightarrow -\infty\\ \phi_2\rightarrow +\infty }}
C_{\mathrm{sing}}^\dagger(\phi_1,\phi_2)\ket{\Psi} = 0~.
\label{eq:Psising}
\end{equation}
On the other hand, the branch wave function for the history corresponding to a
bounce in sLQC is
\begin{equation}
\ket{\Psi_{\text{bounce}}(\phi)}  =  U(\phi-\phi_o)  \lim_{\substack{\phi_1\rightarrow -\infty\\ \phi_2\rightarrow +\infty }} C_{\text{bounce}}^{\dagger} (\phi_1,\phi_2) \ket{\Psi} = |\Psi(\phi)\rangle ~.
\label{eq:Psibounce}
\end{equation}
Thus, the family of histories
$(\text{bounce},\text{singular})$ in sLQC decoheres, 
\begin{equation}\label{eq:dbs}
d(\text{bounce},\text{sing})  =  \bracket{\Psi_{\text{sing}}}{\Psi_{\text{bounce}}} = 0,
\end{equation}
and a bounce is predicted with probability 1,
\begin{subequations}
\begin{eqnarray}
d(\text{bounce},\text{bounce}) & = & \bracket{\Psi_{\text{bounce}}}{\Psi_{\text{bounce}}}
\label{eq:dbb-a}\\
 & = & \bracket{\Psi}{\Psi}
\label{eq:dbb-b}\\
 & = & 1.
\label{eq:dbb-c}
\end{eqnarray}
\label{eq:dbb}%
\end{subequations}
Note that in this analysis, no assumption has been made on the the choice of
state $\ket{\Psi}$, and thus this result holds for all states in the theory.
Thus, we have shown in the consistent histories approach that the bounce is a
universal feature of all states in sLQC. 

We finally note that the existence of bounce at a non-zero volume is tied to
the existence of an upper bound on the expectation values of the energy
density operator of the scalar field: $\langle\hat \rho|_\phi\rangle = \langle
p_\phi\rangle^2/2 \langle V|_\phi\rangle^2$.  For more discussion, see Refs.\
\cite{acs:slqc,dac13a}.  Unlike the Wheeler-DeWitt theory, in sLQC the
spacetime curvature thus never diverges during the evolution.

\section{Discussion}
\label{sec:discuss}

The essence of quantum superposition is that independent reality  
cannot be assigned to the elements of that superposition unless interference
among them vanishes.  In the language of consistent or decoherent histories
``generalized'' quantum mechanics, physical probabilities cannot be inferred
from the transition amplitudes unless the corresponding family of histories is
consistent, as emphasized in \cite{CS10a,CS10b}.  For a closed quantum
system, therefore, it is essential to have available an internally consistent
measure of quantum interference in order to be able to arrive at meaningful
quantum predictions.

In a closed system such as the universe as a whole, an objective measure of
quantum interference is provided by the system's decoherence functional.
Construction of the decoherence functional is therefore an essential
component of any quantum theory of gravity in which one intends to apply the
theory to the whole universe, as in quantum cosmology.

The point of view of the decoherent or consistent histories framework as
applied here%
\footnote{It is to be noted that the broad framework of generalized quantum
theory \cite{hartle91a,lesH,ILSS98a}, quantum measure theory
\cite{sorkin94,*sorkin97a} and related generalizations
does indeed allow for significant generalizations beyond the standard quantum
formalism, and indeed, allowing this freedom -- while retaining characteristic
features of quantum theory such as superposition and interference -- was one
of the theory's originating motivations.  Agreement with standard quantum
theory in familiar regimes does however constrain that freedom considerably.
} %
may be sufficiently unfamiliar to some that it is important to emphasize
that, in almost every respect, it is simply ``quantum mechanics as usual''.
The single -- but crucial -- new concept is the addition of the decoherence
functional to the technical and interpretational apparatus of quantum
theory.  The decoherence functional is essentially an extension of the
concept of quantum state%
\footnote{Especially as viewed from the point of view of the algebraic
formulation of quantum theory \cite{ILS94a,*dac97}.
} %
to provide an objective, internally consistent measure of quantum
interference, thus replacing the vague criterion of measurement by external
classical observers with a rigorously formulated measure that reproduces the
results of classical measurement theory when it is applicable, but extends it
to situations when it is manifestly, and profoundly, \textbf{not} applicable
-- crucially, to closed systems for which the notion of external measurements
is clearly meaningless.  Simple examples of the necessity for such an
extension are easy to come by in quantum gravity.  For example, how is one to
assign probabilities to the quantum density fluctuations that putatively lead
to the large scale cosmological structure we observe today when no classical
systems existed to ``observe'' them?

The focus on ``histories'' may also give the framework an unfamiliar feel.
However, it is precisely for making predictions concerning \emph{sequences} of
quantum outcomes that ordinary measurement-based formulations of quantum
mechanics have no answers, no predictions, for closed quantum systems.  Yet,
patterns of correlations between observable quantities -- paths, or
``histories'' of those observables -- are precisely the kind of quantities in
which one is principally interested in cosmology.  The decoherent histories
framework provides a consistent and rigorous foundation for calculating
quantum probabilities, whether for single quantum events, or sequences
thereof, in terms of quantum amplitudes given by the system's state and
physical inner product.  In a properly formulated generalized quantum theory
of cosmology, the physical meaning of the ``wave function of the universe'' is
unambiguous; there is no need to rely on heuristic arguments
\cite{halliwell91:qcbu} to extract physical predictions
\cite{hartle91a,lesH,CH04,CS10c}.  The methodology for quantum prediction is
precise and clear.

We noted in the introduction that the interpretation of the meaning of
probability in quantum theory quite generally -- not only in the quantum
mechanics of closed systems -- remains controversial.  Notwithstanding, we
are not reluctant to write down expressions for quantum probabilities such as
those found in Sec.\ \ref{sec:apps} with the expectation that their
interpretation is as clear in context as it ever is in quantum mechanics.%
\footnote{We write down and use probabilities for predictive purposes all the
time in physics even when there is no realistic expectation of access to more
than one copy of a system.  In spite of the foundational challenges --
which we fully recognize -- we trust this simple observation is not
controversial.
} %
In recognition nonetheless of the special status of closed systems, we focus
particular attention on quantum predictions which are certain, those for
which the probabilities are 1 or 0 -- or close to it.  For such predictions,
the meaning of the probabilities is unambiguous: the universe either will (or
will not) exhibit the property (or history) in question.  If observation
contradicts this prediction, the theory is simply incorrect.  (Hartle 
\cite{hartle88a,*hartle91a} and Sorkin \cite{sorkin94,*sorkin97a}
in particular have emphasized the special role played by such certain 
predictions in the quantum theory of closed systems.)

In an exactly solvable model of loop quantum cosmology (sLQC) \cite{acs:slqc},
for example, we have shown in a quantum mechanically consistent way that
\emph{all} states in the theory -- whether peaked on classical trajectories at
large volume or not -- ``bounce'' in the sense that the universe must have a
large volume in the limit of large $|\phi|$.  In stark contrast to the
corresponding classical model, and also to its older Wheeler-DeWitt-type
quantization, these quantum universes are not drawn into an infinite density
singularity, and indeed, cannot be.

The role of the large-$|\phi|$ limit in our predictions is worth comment.  As
noted in passing in Sec.\ \ref{sec:bounce}, for a given state $\ket{\Psi}$ it
is certainly possible that the histories described by
$C_{\mathrm{bounce}}(\phi_1,\phi_2)$ and $C_{\mathrm{sing}}(\phi_1,\phi_2)$
will decohere to a high degree of approximation at finite $\phi_{1,2}$.
Indeed, wide classes of states such as states which are semi-classical at late
times, i.e.\ peaked on classical trajectories at large volume, will do so.
However, it is only in the limit that $|\phi|$ becomes large that \emph{all}
states, no matter how ``quantum'' (with no peakedness properties), are
guaranteed to decohere and exhibit a quantum bounce.  This is the role of the
limit: it is in this limit we arrive at a \emph{universal}, and certain,
prediction of sLQC -- including both decoherence and unit probability -- valid
for all states.  The bounce is a robust, universal prediction of sLQC.%
\footnote{The arguments of Refs.\ \cite{fpps12,*pf13a} notwithstanding; see 
our remarks in Sec.\ \ref{sec:bounce} above.
} %

It is of course the case that it was rigorously shown in Ref.\
\cite{acs:slqc,dac13a} that the matter density remains bounded above for all
states in the theory (in the domain of the relevant physical operators.)
Here, we complement this result and add a little more.  First, as emphasized
in Refs.\ \cite{CS10a,CS10b}, the physical question of whether a quantum
universe exhibits a ``bounce'' is fundamentally a prediction about the
correlation of the volume with (at least) two different values of the scalar
field -- at two different emergent `times'.  It is precisely for such
predictions that the question of the decoherence of the corresponding
histories becomes critical.  Here we have shown that the coarse-grained
histories describing a bounce do indeed decohere and predict a bounce with
probability 1.

This actually goes further than the assertion that the matter density remains
bounded above.  In fact, we showed that for arbitrary choice of 
volume $\v^*$, the branch wave function $\ket{\Psi_{\Delta\v^*}}$ describing a
universe with volume $|\v|$ in $\Delta\v^*=[0,\v^*]$ becomes 0 as
$|\phi|\rightarrow\infty$, complementing the result of Ref.\ \cite{acs:slqc}
that the expectation value of the volume becomes infinite in those limits for
all states.
The fact that $\v^*$ is completely arbitrary implies that \emph{all} quantum
states in sLQC will eventually end up at large volume, and therefore, be
described by a superposition of Wheeler-DeWitt quantum states in that regime
\cite{dac13a,dac13c}.  (Of course, that does not mean all states behave
quasiclassically if the state is highly quantum.  It only means that sLQC
passes over to the Wheeler-DeWitt quantum theory at large volume, and that
that limit obtains for every loop quantum state for some range of large
$|\phi|$.  For a detailed analysis of the precise sense in which sLQC
approximates the Wheeler-DeWitt theory, see Ref.\ \cite{acs:slqc}.)  This
rules out, for example, the possibility of a highly quantum state that lingers
indefinitely near the ``big bang'' (i.e.\ at large matter density).

Consistent with the prediction that all states ``bounce'', as noted all states
in sLQC look like a particular symmetric superposition of expanding and
collapsing Wheeler-DeWitt universes at large volume \cite{dac13a}.  While this
is certainly a \emph{necessary} condition for a theory in which a bounce is a
generic feature, one may be led to inquire whether the presence of this
superposition is in some sense the \emph{reason} for the bounce.  The answer
is definitely ``NO''.  In fact, it was shown in Ref.\ \cite{CS10a,CS10b,CS10c}
that a superposition of expanding and contracting universes in the
Wheeler-DeWitt quantization of this same physical model does not, and indeed
\emph{cannot} bounce: all states are sucked in to the singularity at large
$|\phi|$.  If a physical reason for the bounce is to be sought, it is in the
``quantum repulsion'' generated at small volume in loop quantum
gravity,%
\footnote{One way to understand this quantum gravitational repulsion is via
the effective dynamics derived from the effective Hamiltonian constraint.
Modifications to the Friedmann and Raychaudhury equations in effective LQC are
such that they correspond to a repulsive gravity effect at small volumes,
causing the Hubble rate and the spacetime curvature to be bounded
\cite{singh09a}.
} %
and manifested in this model in the ultraviolet cutoff in the dynamical
eigenfunctions \cite{dac13a}, not in the superposition of large expanding and
contracting classical universes.

We have so far exhibited the construction of a generalized
decoherent-histories quantum theory in two mathematically complete
quantizations of cosmological models, sLQC in this analysis, and the
corresponding Wheeler-DeWitt model in an earlier work
\cite{CS10a,CS10b,CS10c}, demonstrating how these theories may be used to
arrive at quantum mechanically consistent physical predictions.
%
In both of these models, the presence of an internal time, a monotonic
physical clock provided by the scalar field in the canonical quantization of
the models, simplified the construction of the generalized quantum theories
through the structural analogy with ordinary quantum particle mechanics.  Is
this feature essential to the construction of a generalized quantum theory?
The answer is no.  A path integral formulation for a closed Bianchi model was
studied in detail in Ref.\ \cite{CH04} and references cited therein, and upon
which much of the present work is based.  More rigorously, we have now
constructed along the same lines the
generalized quantum theory  
for a path-integral (spin foam) quantization \cite{ach09,ach10a,ach10b} of
the same physical model (flat scalar FRW) as studied here \cite{CS12c}.  This
will again be employed to study the same physical questions examined here, in
particular, the quantum behavior near the classical singularity.  This
example shows that the presence of an internal time in the model, while
convenient, is not essential to the formulation of its class operators,
branch wave functions, and decoherence functional.%
\footnote{The most significant difference between canonical and path integral
formulations of generalized quantum theories concerns the manner in which
quantum alternatives are specified.  In canonical constructions one is largely
limited to histories that can be specified by superpositions of chains of
projections onto ranges of physical observables.  In theories defined by a
sum-over-histories, alternatives can be specified by partitioning the paths
that contribute to the sum according to whether the paths exhibit some
particular property, and branch wave functions by restricting to integrals
(sums) over the appropriate class of paths.  This can lead to interesting
differences between canonical and path integral formulations, even in ordinary
non-relativistic quantum mechanics.
} %
Taken together, these examples lay the foundation for the construction and
application of generalized quantum theories -- the quantum theory of closed
physical systems -- in quantum cosmology and quantum gravity more broadly.

\appendix* 




\begin{acknowledgments}

Research at the Perimeter Institute is supported by the Government of Canada
through Industry Canada and by the Province of Ontario through the Ministry of
Research and Innovation.  D.C.\ would like to thank the Department of Physics
and Astronomy at Louisiana State University, where portions of this work were
completed, for its hospitality.  D.C.\ was supported in part by a grant from
FQXi.  P.S.\ is supported by NSF grant PHYS1068743.

\end{acknowledgments}


\ifthenelse{\arxiv=1}{%
\bibliography{dLQC}
}{%
\bibliography{global_macros,../Bibliographies/master}%
}%

\end{document}